# One Stone, Two Birds:
# Using Vapor Kinetic Energy to Detect and Understand Atmospheric Rivers

## Author Information

### Affiliations

Hing Ong[1]* & Da Yang[2]

[1]Argonne National Laboratory, Lemont, IL, USA

[2]University of Chicago

*This work was done when H.O. was a postdoctoral scholar in D.Y.'s group at University of California, Davis.

### Corresponding author

Correspondence to: Da Yang (e-mail: dayang@uchicago.edu)

### ORCID

Hing Ong (0000-0001-9827-9318)

Da Yang (0000-0001-7180-6827)

Supplementary Videos at https://www.yang-climate-group.org/publications.

**Abstract**

Poleward water vapor transport in the midlatitudes mainly occurs in meandering filaments of intense water vapor transport, spanning thousands of kilometers long and hundreds of kilometers wide and drifting eastward. The water vapor filaments are known as atmospheric rivers (ARs). They can cause extreme wind gusts, intense precipitation, and flooding along densely populated coastal regions. Many recent studies about ARs focused on the statistical analyses of ARs, but a process-level understanding of ARs remains elusive. Here we show that ARs are streams of air with enhanced vapor kinetic energy (VKE) and derive a governing equation for Integrated VKE (IVKE) to understand what contributes to the evolution of ARs. We find that ARs grow mainly because of potential energy conversion to kinetic energy, decay largely owing to condensation and turbulence, and the eastward drift is primarily due to horizontal advection of VKE. Our VKE framework complements the integrated vapor transport framework, which is popular for identifying ARs but lacks a prognostic equation for understanding the physical processes.

An atmospheric river (AR) is a narrow corridor of running water vapor pushed along by strong winds at low altitudes—a low-level jet[1]. ARs often form with the development of extratropical cyclones and their associated cold fronts[1-3]. On average, each AR transports more than the flow of the Amazon River[4,5]. ARs contribute most of the poleward water vapor transport in the midlatitudes[5,6] and can lead to extreme winds[7] and precipitation in coastal regions around the world[7-13]. For example, in October 2021, an atmospheric river made landfall in California and brought strong winds with peak gusts around 70 mph and historic heavy rainfall, leading to flooding, power outages, and mud slides[14]. ARs' impact on hydrological cycle has aroused growing research interest in the past decade[15]. The scientific community has collaborated to detect ARs and quantify their statistical features, including frequency, intensity, duration, *etc.* in the AR Tracking Method Intercomparison Project (ARTMIP)[6,16,17]. However, it remains unclear what physical process provide energy to maintain ARs, and what physical processes lead to the eastward drift of ARs.

Since ARs are associated with both high humidity and strong winds, ARs are often detected using integrated vapor transport (IVT)[6]:

$$\text{IVT} \equiv \sqrt{\left(-\frac{1}{g}\int_{p_\text{B}}^{p_\text{T}} qu\text{d}p\right)^2 + \left(-\frac{1}{g}\int_{p_\text{B}}^{p_\text{T}} qv\text{d}p\right)^2}, \qquad (1)$$

where the variables are defined as follows: $q$, specific humidity; $u$, zonal velocity; $v$, meridional velocity; $g$, gravity; and $p$, pressure. The lower limit, $p_\text{B}$, is surface pressure. The upper limit, $p_\text{T}$, is a constant pressure beyond which specific humidity is negligible; $p_\text{T}$ is set to 200 hPa[16]. ARs' IVT anomalies are sometimes dominated by high humidity and sometimes by strong winds[18].

Unfortunately, it is challenging to derive a budget equation or conservation law for IVT, so most previous studies analyzed the column water vapor (CWV) budget to study the evolution of ARs[19-23]. They showed that horizontal convergence is the primary source of ARs' water vapor, which is then lost through water vapor condensation and precipitation[19-22]. This result broadly agrees with a Lagrangian analysis of CWV anomalies[23]. These studies implicitly assumed that evolution of winds is of secondary importance to the physics of ARs, a compromise of lacking an analysis method that accounts for both winds and humidity. However, since the development of ARs is closely tied to the lifecycle of baroclinic waves and extratropical cyclones[1-3], and their impact is largely attributed to intense winds, it is imperative to establish an analysis framework that organically incorporates both wind and humidity elements.

Here we propose using vapor kinetic energy (VKE) to study ARs. Starting from vapor transport, $qu$ and $qv$, we define VKE as $[(qu)^2+(qv)^2]/2 = q^2K$, where $K$ denotes KE in horizontal winds per unit mass ($[u^2+v^2]/2$). If we consider vapor transport as specific humidity-weighted momentum, then VKE is the corresponding energy. By construction, VKE always increases with the magnitude of vapor transport. Then, we define integrated VKE (IVKE) as follows:

$$\text{IVKE} \equiv -\frac{1}{g}\int_{p_\text{B}}^{p_\text{T}} q^2 K \, dp. \qquad (2)$$

We will show that IVKE is as effective as IVT in AR detection and that the budget of IVKE exposes the physics of ARs' evolution.

We first show the IVT- and IVKE-based detection results using the TempestExtremes algorithm[24] and the MERRA-2 data. The detection results are almost identical in terms of AR presence (Video S1) and frequency (Figure 1a, b). The frequency difference between IVT and IVKE is on the order of 1 day per year (Figure 1c). Both Video S1 and Figures 1a, b seamlessly juxtapose the panels of results using IVT and IVKE. The results match in most details, and the transition between the results is almost continuous; the detected ARs on the right of the left panel (using IVT) appear on the left of the right panel (using IVKE) as if the right panel was an extension of the left panel. Also, the AR frequency in Figure 1 is similar to the ensemble median of ARTMIP in terms of the spatial patterns and the maximum values[17]. We then use an independent detection algorithm developed by Mundhenk et al.[25] to further compare the IVT- and IVKE-based AR detection. The overall results remain similar: using IVKE can effectively capture AR events (Figure 1d-f). These results suggest that ARs are streams of air with enhanced IVKE, and the evolution of IVKE may explain the evolution of ARs.

Figure 1c, f provides an opportunity to compare AR detection results based on different algorithms. The frequency difference due to switching from TempestExtremes to the Mundhenk et al. algorithm is on the order of 10 days per years, which is similar to the ensemble interquartile range of ARTMIP[17]. Clearly, switching from IVT to IVKE introduces less AR frequency difference than switching between detection algorithms.

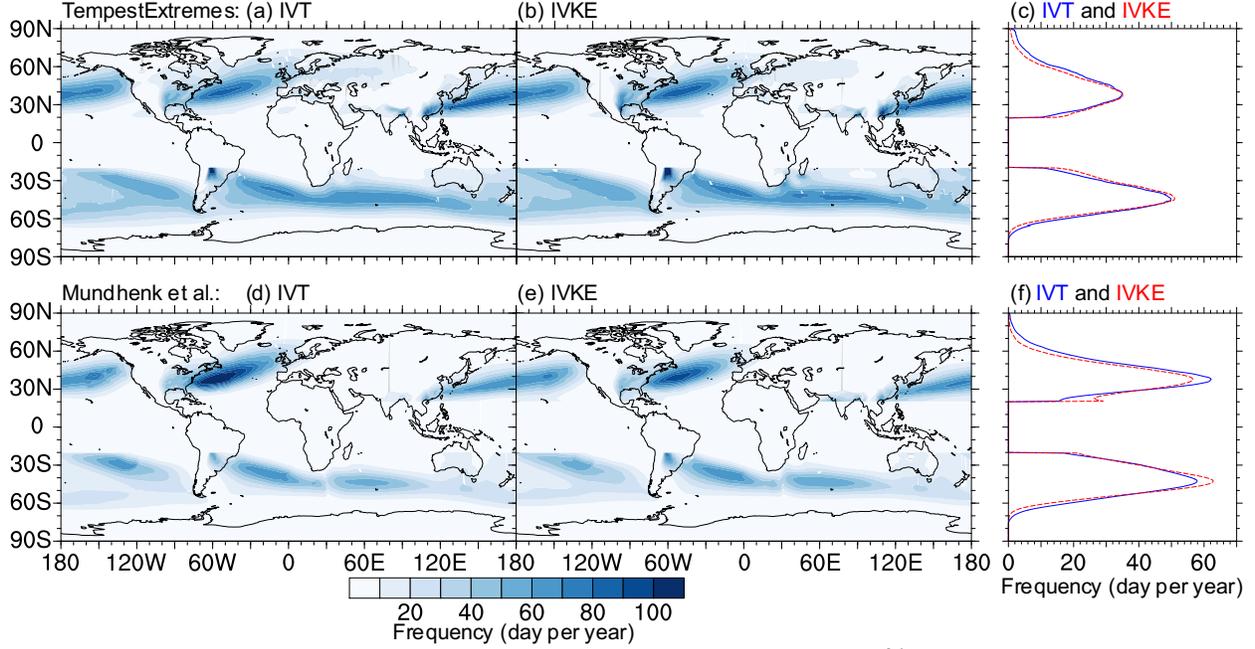

**Figure 1.** AR frequency using the algorithms of TempestExtremes[24] (upper) and Mundhenk et al.[25] from MERRA-2 data (1980 to 2019). Panels (a)(d) use IVT, and (b)(e) use IVKE. Panels (c)(f) show zonally averaged results using IVT (blue solid) and IVKE (red dash).

We then derive a budget equation for IVKE by combining the conservation of momentum and water vapor to understand ARs' evolution (Supplementary Materials: Equations S1-S7). The IVKE tendency can be decomposed into two components: one component is the sum of the vertical integral of VKE tendency, and the other component is related to chances in surface pressure with time (Equation S6-S7), where the latter term is negligible in our analysis. Therefore, the key equation is the VKE tendency equation (simplified from Equation S5):

$$\frac{\partial q^2 K}{\partial t} = -\mathbf{u} \cdot \mathbf{\nabla}_p (q^2 K) - \omega \frac{\partial q^2 K}{\partial p} - q^2 \mathbf{u} \cdot \mathbf{\nabla}_p \Phi + q^2 \mathbf{u} \cdot \mathbf{F_T} + 2KqS_\mathrm{M} + \text{Other}. \qquad (3)$$

The right-hand side of Equation 3 shows that the primary contributors to the evolution of VKE are horizontal advection of VKE ($-\mathbf{u} \cdot \mathbf{\nabla}_p(q^2 K)$), vertical advection of VKE ($-\omega \frac{\partial q^2 K}{\partial p}$), potential energy (PE) conversion to KE ($-q^2 \mathbf{u} \cdot \mathbf{\nabla}_p \Phi$), turbulent dissipation of KE ($q^2 \mathbf{u} \cdot \mathbf{F_T}$, including surface flux of momentum), and condensation of vapor ($2KqS_\mathrm{M}$; MERRA-2 combines subgrid-scale moist convection, condensation, and evaporation of condensate into one variable, $S_M$[26]). PE conversion to KE can be interpreted as flowing down a slope of an isobaric surface ($-\mathbf{u}\cdot\nabla_p\Phi$)[27]. This term is absent from the CWV budget—widely used in earlier AR studies—but turns out to be of leading order importance in our following analysis. Latent heat release associated with vapor condensation cannot directly change VKE but may affect VKE through enhancing PE conversion to KE[28]. The other terms are negligible in our analysis (See Equations S5-S10 and Table S1, where the analysis tendency introduced by the reanalysis[29] is included, and turbulence tendency of vapor includes surface flux of vapor[26]).

We make a composite AR and analyze its IVKE budget using MERRA-2 reanalysis data[29] from 2010 to 2019. Our composite AR is based on a widely used composite method to study extratropical cyclones[30], baroclinic waves[31,32], and convectively coupled circulations[33,34]. This analysis focuses on an active AR region in the North Pacific, where the IVKE temporal variance maximizes among regions that do not have tropical cyclones (Figure S1). We first calculate the time series of areal mean IVKE in the 1°×1° box centered at 155°E 35°N and calculate its anomaly from the temporal mean. We then perform a regression analysis at each grid point: we regress anomalous fields of VKE, IVKE, and their tendency terms on the time series of IVKE anomaly that we just obtained. We multiply the regression coefficient of a given field at each grid point by one sample standard deviation of the IVKE time series to reconstruct the corresponding field of the composite AR. See the Methods Section C for more details about the composite.

Through the lens of IVKE, the composite AR is an elongated ellipse with the major axis pointing toward the east-northeast (contours in Figure 2). The total IVKE tendency leads the AR's IVKE anomaly by about a quarter cycle (Figure 2a), suggesting that the total tendency makes little change to the IVKE amplitude but promotes the AR's eastward drift. We find that the horizonal VKE advection (Figure 2b) has similar spatial structure and even overall amplitude to those of the total IVKE tendency (Figure 2a). This result shows that the horizontal advection is the primary drifting mechanism of the AR. We further decompose the horizontal advection of VKE into vapor and KE parts, and the vapor part leads to an eastward drift (Figure S2a) while the KE part may lead to an east-northeastward drift (Figure S2b). This result adds to earlier moisture budget analyses that only considered the role of horizontal advection of moisture.

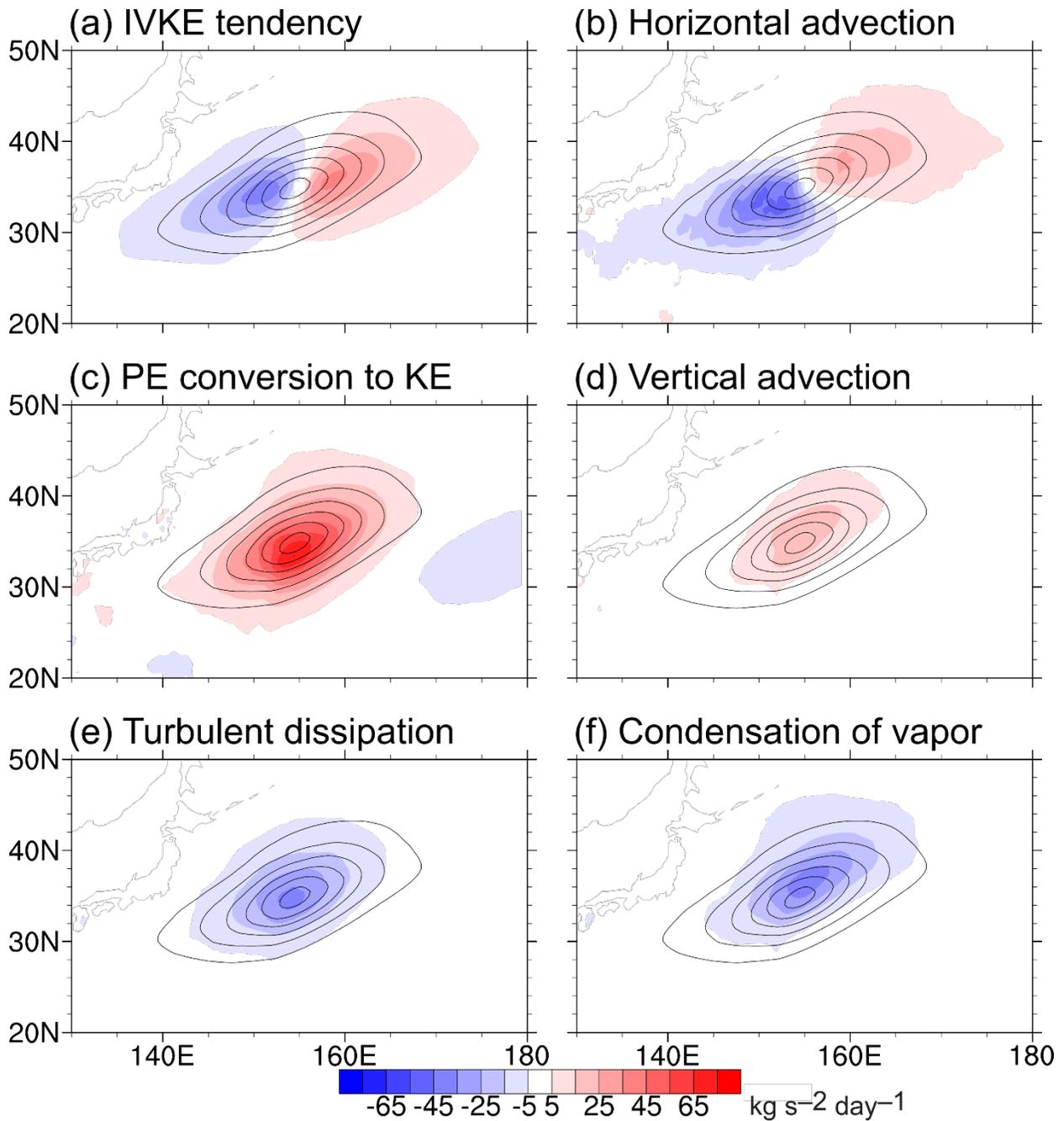

**Figure 2.** Plan view of the AR composite anomalies by regressing upon areal mean IVKE in the 1°×1° box centered at 155°E 35°N from MERRA-2 data (2010 to 2019). The contours are IVKE (interval: 6 kg s$^{-2}$). The shades are IVKE tendencies due to (a) total, (b) horizontal advection of VKE, (c) PE conversion to KE, (d) vertical advection of VKE, (e) turbulent dissipation of KE, and (f) condensation of vapor.

The IVKE tendency due to PE conversion to KE is in phase with the AR's IVKE anomaly (Figure 2c). Therefore, the energy conversion increases the AR's IVKE anomaly and is the primary process responsible for the AR's development and maintenance. IVKE tendency that lies outside of the AR's IVKE anomaly does not project onto the AR's IVKE anomaly and, thus, cannot affect the AR's intensity. This PE-to-KE conversion is essential to maintain the low-level jets[28], and this process was overlooked in earlier moisture budget analyses. We emphasize that the PE-to-KE conversion mainly arises from the anomalous winds that can have a significant ageostrophic component. For example, Figure S3a shows specific humidity, geopotential, and wind anomaly of the composite AR at 950 hPa, where VKE maximizes. There is a significant anomalous wind component that slides down the geopotential gradient, converting PE to KE and accelerating the low-level jet (Figure S3a). On the other hand, the mean winds mainly follow the geopotential contours and contribute less to the PE-to-KE conversion (Figure S3b). We see similar patterns throughout pressure levels from 975 to 850 hPa (Figure S4 and Video S2).

The vertical VKE advection also increases AR's IVKE anomaly near the center of the AR, where IVKE is rich (Figure 2d), so the vertical advection also contributes to the AR's development and maintenance. On the other hand, the AR loses KE through turbulent dissipation (Figure 2e) and loses vapor through condensation (Figure 2f). Both processes contribute to the decay of the AR. Other physical processes make negligible contributions to the IVKE budget and show no signal in plots using the color scheme in Figure 2. To test the robustness of our results, we make composite ARs in the AR active regions in the North Atlantic (Figure S5) and South Atlantic (Figure S6) and analyze their IVKE budget. The overall results remain similar. See Methods Section C for more details about robustness testing.

Our analysis framework resolves the vertical dimension of the composite AR. Figure 3 shows an east-west vertical cross section through the center of the AR along 35°N. The VKE anomaly increases from the bottom to 950 hPa due to the strong low-level jet and high humidity. Then VKE decreases with altitude mainly due to the exponential decrease of water vapor (Figure 3). The vertically resolved results supports the map view analysis (Figure 2). The total VKE tendency promotes the AR's eastward drift (Figure 3a). The horizonal VKE advection (Figure 3b) shows a similar pattern and magnitude to the total VKE tendency (Figure 3a). We also decompose the horizontal advection of VKE into vapor and KE parts (Figure S7a, b), and the results are consistent with the map view (Figure S2a, b). The PE conversion to KE is a source of VKE (Figure 3c), and the turbulent dissipation of KE is a sink (Figure 3e). Both processes show bottom-heavy structures. Condensation contributes to the AR's decay and is distributed from 900 hPa to 500 hPa (Figure 3f). The vertical VKE advection increases VKE above 950 hPa but decreases VKE from the bottom to 950 hPa (Figure 3d). Since the AR's VKE anomaly maximizes at 950 hPa, large-scale ascent in the AR advects VKE-rich air from lower altitude to higher above 950 hPa but the opposite below. When we decompose the vertical advection, the vapor part increases VKE (Figures S2c and S7c) while the KE part decreases VKE (Figures S2c and S7c) because specific humidity is bottom-heavy while KE is top-heavy (Video S2). Figure S8 shows a south-north vertical cross section through the center of the AR along 155°E. The AR's anomalous air motion is poleward and upward (Figure S8). The VKE contours above 950 hPa resemble a hill where the poleward slope is steeper than the equatorward slope (Figure S8). Such motion and asymmetry are consistent with previous studies and are associated with the structure of the cold front[1,4,19].

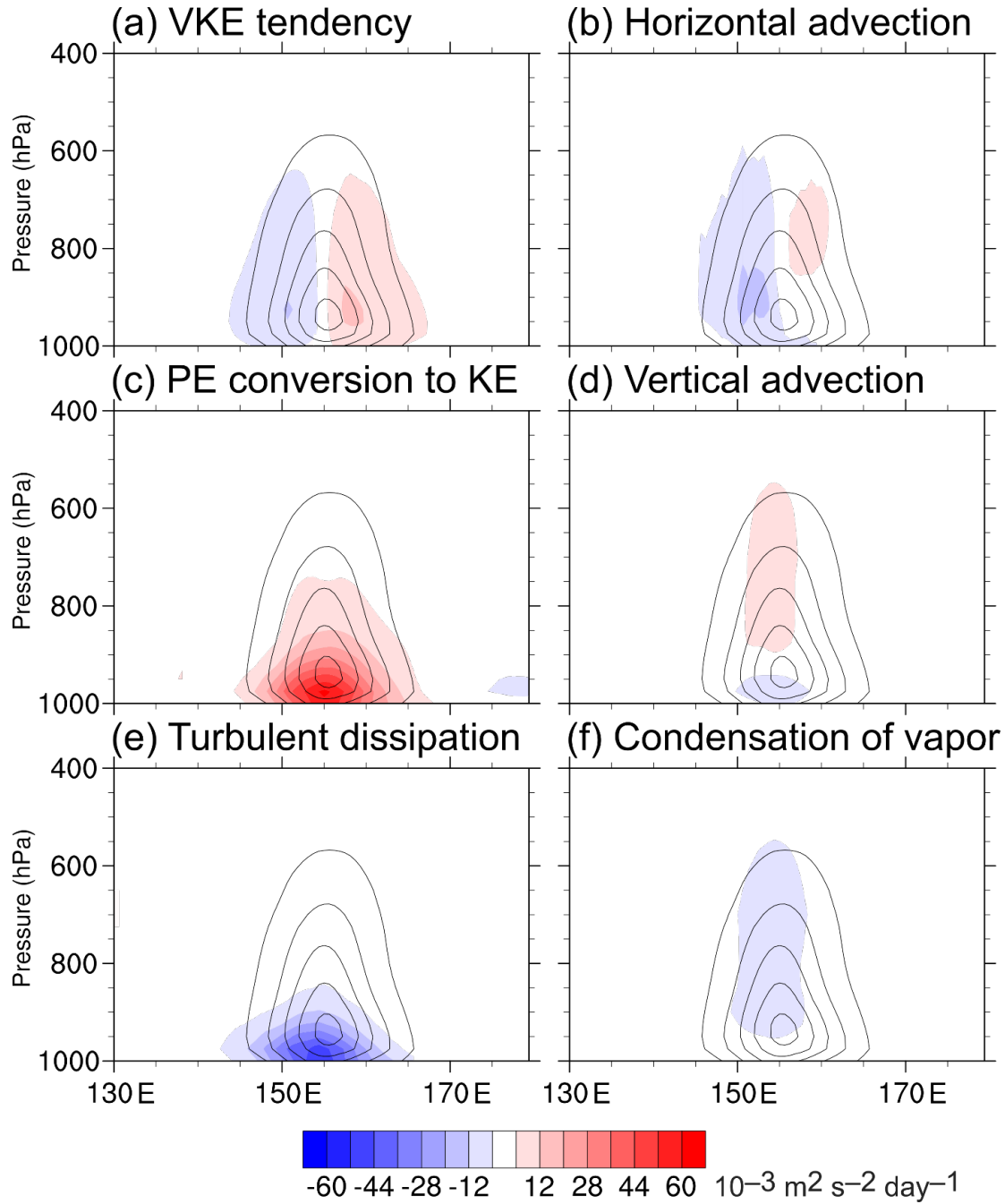

**Figure 3.** Like Figure 2 but vertical cross section along 35°N of the AR composite anomalies (average taken between 34.5°N and 35.5°N). The contours are VKE (interval: $2 \times 10^{-3}$ m$^2$ s$^{-2}$). The shades are VKE tendencies due to (a) total, (b) horizontal advection of VKE, (c) PE conversion to KE, (d) vertical advection of VKE, (e) turbulent dissipation of KE, and (f) condensation of vapor.

We project each tendency term in Equation S7 onto the regressed field of IVKE anomaly to quantify the contribution of the physical processes to the growth or decay of the composite AR (see Methods Section A and Equation 4a). The projection value has a unit of inverse timescale and can be interpreted as a rate of growth or decay. We can also interpret this projection process as calculating the spatial correlation between a specific IVKE tendency term and the IVKE anomaly, given that the IVKE tendency terms can reinforce the IVKE anomaly or AR intensity only if they are in phase. On the other hand, if an IVKE tendency term has significant values outside the AR's IVKE anomaly, this physical process would not change the AR intensity. This quantitative analysis (Figure 4a) agrees with our qualitative analysis (Figure 2). The IVKE budget is well closed given a residual of only 0.6% of the sum of the IVKE sources (Table S1).

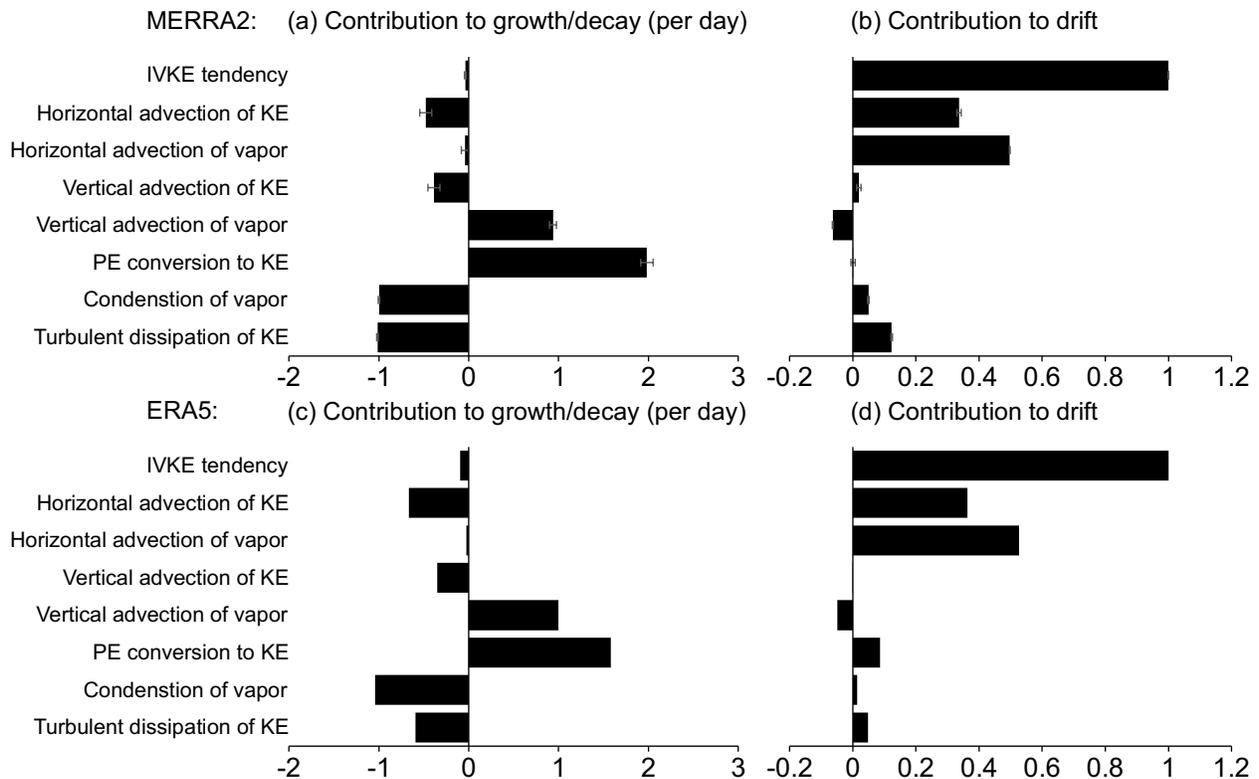

**Figure 4**. Contribution of physical processes to the growth or decay (left) and drift (right) of the composite AR in Figure 2 using MERRA-2 (upper) and ERA5 (lower) data (2010 to 2019) within 130°E to 180°, 20°N to 50°N. In the upper panels, all the terms are directly calculated from MERRA-2 data, and the residuals (Equations S8 to S10) are represented by the error bars. In the lower panels, condensation of vapor and turbulent dissipation of KE are not provided by ERA5 but are instead estimated from Equations S11 and S12. See Supplementary Text for details.

We then measure the contribution to the AR's eastward drift by projecting each term on the right-hand side of Equation (3) onto the total IVKE tendency of the composite AR (See Methods Section A and Equation 4b). The horizontal VKE advection is the primary mechanism for the AR's eastward drift (Figure 4bd and Table S1). We further show that the horizontal advection of vapor contributes more to the eastward drift than the horizontal advection of KE (Table S1). Moving the boundary of the projection area does not affect the results unless cutting any contoured region of the AR's IVKE anomaly (Figure 2) or shaded region of the AR's IVKE total tendency anomaly (Figure 2a). The projection boundary cuts a region of negative PE-to-KE conversion (Figure 2c), but our calculations show that changing the cut does not affect the results, as the negative PE-to-KE region does not overlap with the area of significant AR's IVKE or its tendency.

To test the robustness of our composite and IVKE analysis, we perform identical calculations using an independent dataset, the ERA5 reanalysis data[35], from 2010 to 2019. The ERA5 results (Figures 4cd and S9) agree well with the MERRA-2 results (Figures 2 and 4ab). The minor discrepancies observed might be attributed to variations in the sampling frequency: ERA5 saves instantaneous data every hour, while MERRA-2 saves time-averaged data every 3 hours. This agreement between the two datasets suggests that our analysis methods are reliable, and the emerging physics is robust regardless of the dataset employed.

The above analyses primarily focus on composite ARs at their peak intensity. What are the roles of different physical processes in different stages of ARs' lifecycles, including intensification and decay? We present a case study toward addressing this question. A historic AR made landfall onto the west coast of North America during 12 to 15 UTC, 4 January 2023. This AR intensified before its landfall and decayed after its landfall. Video S3 shows the 3-hourly evolution of this AR from 18 to 21 UTC, 3 January 2023 to 18 to 21 UTC, 4 January 2023. We take the projection of IVKE tendency ($\partial$IVKE/$\partial$t) on IVKE (Equation 4a) as the rate of growth (positive) or decay (negative); our projection region is 160°W to 100°W, 20°N to 50°N. The growth rate was 0.91±0.02 day$^{-1}$ from 18 to 21 UTC, 3 January 2023 (Figure S10), so the AR intensity changed by 12% within 3 hours. The AR intensity kept increasing by 10% in the next 3 hours (Video S3). Then, the AR maintained and slightly increased its intensity from 00 to 15 UTC, 4 January 2023 with a growth rate fluctuating from –0.04 to 0.40 day$^{-1}$ (Video S3). The AR decayed after landfall (Video S3). The growth rate was –0.75±0.01 day$^{-1}$ from 18 to 21 UTC, 4 January 2023 (Figure S11), so the AR intensity changed by 9% within 3 hours. Throughout the AR evolution, the last four terms on our growth rate diagrams (Figure 4ac, S10b, S11b, and Video S3), including the contributions of PE-to-KE conversion, vertical advection of vapor, turbulent dissipation of KE, and condensation of vapor, remain as the most prominent terms, qualitatively similar to those in the composite AR (comparing Figures S10b and S11b and Video S3 to Figure 4ac). In addition, though the partition between the horizontal advection of KE and vapor changes throughout the AR lifecycle, their sum remains negative (Video S3). In general, the IVKE sources are in excess during the growth stage (Figure S10), while the IVKE sinks are excessively negative during the decay stage (Figure S11).

Here we propose to use VKE to detect and study ARs, killing two birds with one stone. IVKE is as effective as IVT in identifying and monitoring ARs. Additionally, the budget of IVKE distills fundamental AR physics: PE conversion to KE is the primary energy source to sustain ARs and is balanced by turbulent dissipation and condensation; horizontal advection of VKE leads to the eastward drift of ARs. Conversion from PE to KE is the main energy source for the low-level jet[28]. Therefore, our IVKE results suggest that the low-level jet is an integral part of ARs. Additionally, the advection of both vapor and KE contributes to ARs' eastward drift, indicating that both humidity and winds are important to ARs' evolution. This analysis framework complements earlier studies that focused on either ARs' humidity evolution or the role of extratropical cyclones in ARs' development, which puts AR research in a firmer footing in atmospheric dynamics.

One may think that analyzing the VKE budget is equivalent to analyzing the budgets of specific humidity and KE separately. However, that is not true. Due to stronger winds aloft, KE budget would emphasize much on the energetics in the upper troposphere, whereas lower-troposphere winds are most relevant to water vapor transport and thereby AR dynamics. VKE is the KE weighted by $q^2$. It shows a bottom-heavy structure and allows us to focus on the dynamics associated with low-level jets that effectively transport water vapor.

Unresolved physics, such as boundary layer turbulence and atmospheric convection, contribute substantially to the IVKE budget. This result suggests that although ARs span hundreds or even thousands of kilometers, accurately simulating and forecasting ARs require realistic representations of these sub-grid scale physical processes.

What to do next? ARs have different flavors and can occur at various locations, causing distinct extreme conditions. For example, ARs also occur in the eastern Pacific and often make landfalls over the west coast of North America. Additionally, there are "windy" and "wet" ARs, producing different surface winds and precipitation[18]. Does the dominant balance of the VKE budget remain the same for all AR flavors at various locations? In this paper, we primarily focus on analyses of composite ARs that can be considered as a developed system. What are the roles of different physical processes in different stages of ARs' lifecycles, e.g., the generation, intensification, and decay? What leads to a windy AR versus a wet AR? ARs become weaker under the influence of industrial aerosols but become stronger with increasing greenhouse gases[36]. What leads to the distinct responses of ARs? Our IVKE framework provides a quantitative analysis framework to address these pressing questions that are key to preparing for AR-induced hydrologic extremes.

# Methods

## A. Numerical analyses of IVT, VKE, IVKE, and tendency terms

We calculate IVT, VKE, IVKE, and all tendency terms (all terms in Equation S5 and S7) from the Modern-Era Retrospective analysis for Research and Application, version 2 (MERRA-2)[29]. To calculate IVT, VKE and its tendency, IVKE and its tendency, and surface pressure tendency effect, we take the surface pressure, specific humidity, eastward wind, and northward wind from the 3-dimensional, 3-hourly, instantaneous, pressure-level, assimilated meteorological fields (inst3_3d_asm_Np in MERRA-2)[37] from 1980 to 2020. We use the Boer scheme[38,39] for the vertical integration and forward differencing for VKE and IVKE tendencies and surface pressure tendency effect. Note that the instantaneous time steps and the tendency time steps are staggered. In addition, taking the layer heights and vertical pressure velocity from the 3-dimensional, 3-hourly, time-averaged, model-level, assimilated meteorological fields (tavg3_3d_asm_Nv[40]) from 2010 to 2019, we calculate VKE and IVKE tendencies due to PE conversion to KE and horizontal and vertical advection of both KE and vapor using central differencing in space and linearly interpolate them from the model levels to the pressure levels. We also calculate the abovementioned terms from the ERA5 global reanalysis[35] from 2010 to 2019, which are 3-dimensional, 1-hourly, instantaneous, pressure-level, assimilated meteorological fields. To calculate all the other tendency terms, we further take the corresponding time-averaged wind tendencies (all variables in tavg3_3d_udt_Np in MERRA-2)[41] and moisture tendencies (all variables starting with "DQVDT" in tavg3_3d_qdt_Np in MERRA-2)[42] from 2010 to 2019. In the wind tendency data, horizontal advection, vertical advection, Coriolis effect, curvature effect, and geopotential gradient are combined into a tendency of winds due to dynamics. Also, in the moisture tendency data, horizontal advection and vertical advection are combined into a tendency of specific humidity due to dynamics. We take these wind and moisture tendencies due to dynamics to calculate VKE and IVKE tendencies due to dynamic tendency of KE and vapor. We denote any given term on the right-hand side of Equation (3) as $X$. Then following Anderson and Kuang[33]'s method and notation: we estimate $X$'s contribution to AR's growth and drift by projecting $X$ onto IVKE and total IVKE tendency ($\partial_t$IVKE):

contribution to growth or decay = $\|X(IVKE)\|/\|(IVKE)^2\|$, and (4a)

contribution to drift = $\|X\partial_t IVKE\|/\|(\partial_t IVKE)^2\|$. (4b)

See Supplementary Text and Equations for details and a further decomposition of PE conversion to KE.

### B. AR detection and frequency

We detect ARs with two algorithms. First, we use the TempestExtremes[24] algorithm with thresholds following McClenny, Ullrich, and Grotjahn[43] except following Rhoades et al.[44] for omitting the additional threshold filtering out tropical cyclones. This method features the Laplacian of IVT < –40,000 kg m$^{-1}$ s$^{-1}$ rad$^{-2}$. For comparison, we test this method with the Laplacian of IVKE instead of IVT with various threshold values. Among the threshold values, we select the Laplacian of IVKE < –2,000 kg s$^{-2}$ rad$^{-2}$, which best reproduces the globally averaged AR frequency using IVT with the above threshold (IVT < –40,000 kg m$^{-1}$ s$^{-1}$ rad$^{-2}$); the AR frequency is the ratio of time steps with AR presence to total. All other thresholds remain the same, including the exclusion of grid points between 20°S and 20°N and the exclusion of blobs with less than 50 connected grid points.

We also detect ARs using the algorithm by Mundhenk, Barnes, and Maloney[25]. This method uses the following thresholds: minimum IVT = 250 kg m$^{-1}$ s$^{-1}$, minimum blob size = 150 contiguous grid points, minimum length-width ratio = 1.6, minimum orientation off the parallel = 0.95 rad, minimum length = 25 grid points, minimum mean IVT anomaly = 305 kg m$^{-1}$ s$^{-1}$, and minimum

eccentricity = 0.87. We exclude grid points between 20°S and 20°N. When switching to IVKE, we test various minimum IVKE and minimum mean IVKE anomaly and find the combination that best reproduces the globally averaged AR frequency using Mundhenk, Barnes, and Maloney's thresholds. Optimal results are found when the minimum IVKE = 14 kg s$^{-2}$ and the minimum mean IVKE anomaly = 20 kg s$^{-2}$. We use MERRA-2 data from 1980 to 2020 for both algorithms.

C. AR composite and IVKE budget analysis

We design an AR composite based on IVKE temporal variance independently from any AR detection method. Here we use data from 2010 to 2019. We sought the IVKE standard deviation map (Figure S1) for local maxima of IVKE variance without overlapping with tropical cyclone tracks[45]. We select three spatial maxima of IVKE variance at 155°E 35°N (North Pacific), 60°W 40°N (North Atlantic), and 27.5°W 37.5°S (South Atlantic). The corresponding plotting and projecting regions are 130°E to 180°, 20°N to 50°N (North Pacific), 85°W to 35°W, 25°N to 55°N (North Atlantic), and 52.5°W to 2.5°W, 52.5°S to 22.5°S (South Atlantic). We confirm that there is high AR frequency at the selected points (Figure 1) and no tropical cyclone track within 2° around the selected points from 2010 to 2019[45]. We take the time series of spatial averaged IVKE in the 1°×1° box centered at the selected points as an AR index (e.g., Figure S12). Also, we removed the temporal mean from the fields of the VKE, IVKE, and all the tendency terms as anomalies. Then, we regress the anomaly at each grid point upon the AR index and multiply the regression coefficient at each grid point by one sample standard deviation of the AR index. This calculation yields the composite AR. We test the robustness of our composite results by switching between MERRA-2 and ERA5 data. Additionally, we excluded the time steps when AR is not detected in the 1°×1° box centered at 155°E 35°N using the TempestExtremes as described in Part B of Methods, and the results remain similar (Comparing Figures 2 to S13). All shown regressed fields are significant at 95% confidence level at each grid point with the two-tailed Student's *t*-test, where the degrees of freedom account for lag autocorrelation[39,46]. The statistically insignificant data points are small valued, so we color them in white in the figures. We analyze IVKE budget of the composite AR by projection as described in Part A of Methods. We test the robustness of our projection results by varying the boundary of our composite domain, and the results remain similar as long as the boundary does not cut off any composite AR signal (Comparing Figures 4ab to S14).

## Data Availability

The MERRA-2 data are available in Goddard Earth Sciences Data and Information Services Center (GES DISC) at doi:10.5067/QBZ6MG944HW0[37], doi:10.5067/CWV0G3PPPWFW[41], and doi:10.5067/A9KWADY78YHQ[42]. The ERA5 data are available in Research Data Archive at the National Center for Atmospheric Research at doi:10.5065/BH6N-5N20[47]. The tropical cyclone track aiding the AR regression region selection are visualized at https://coast.noaa.gov/hurricanes/ while the data are available in NOAA National Centers for Environmental Information at doi:10.25921/82ty-9e16[48].

## Code Availability

The TempestExtremes version 2.2 algorithm for the AR detection is available in GitHub at https://github.com/ClimateGlobalChange/tempestextremes. The AR detection algorithm of Mundhenk, Barnes, and Maloney is available in Digital Collections of Colorado at http://hdl.handle.net/10217/170619. The NCAR Command Language (NCL) version 6.5.0 built-in functions for the numerical analyses are available at doi:10.5065/D6WD3XH5[39]. Custom scripts for applying the NCL built-in functions are available upon request.

# Acknowledgements

This research is supported by a Packard Fellowship awarded to D.Y. We thank Paul Ullrich for the technical support on the use of TempestExtremes algorithm.

# Author contributions

D.Y. designed and oversaw the research. H.O. performed the research. H.O and D.Y. wrote the paper together.

# Competing interests

The authors declare no competing interests.

Supplementary Information for

# One Stone, Two Birds:

# Using Vapor Kinetic Energy to Detect and Understand Atmospheric Rivers

Hing Ong[1] & Da Yang[2]

[1]Argonne National Laboratory, Lemont, IL, USA

[2]University of Chicago

**Contents of this file**

Supplementary Text and Equations S1 to S13

Supplementary Figures S1 to S15

Supplementary Table S1

**Additional Supporting Information (Files uploaded separately)**

Supplementary Video S1. AR presence (blue) detected using the TempestExtremes algorithm with IVT (left panel) and IVKE (right panel) from 6-hourly MERRA-2 data from 00 UTC 1 Jan. 2019 to 18 UTC 31 Jan 2019.

Supplementary Video S2. The specific humidity, geopotential height, and winds of the composite AR on pressure levels from 975 to 850 hPa every 25 hPa. The plotting convention is the same as Figures S3 and S4.

Supplementary Video S3. The evolution of an AR event from 18 to 21 UTC, 3 January 2023 to 18 to 21 UTC, 4 January 2023 from 3-hourly MERRA-2 data. The plotting convention is the same as Figures S10 and S11.

**Supplementary Text and Equations**

*Derivation of the IVKE budget*

To derive the VKE prognostic equation, we start from the moisture prognostic equation:

$$\frac{\partial q}{\partial t} = -\mathbf{u} \cdot \nabla_p q - \omega \frac{\partial q}{\partial p} + S_M + S_T + S_C, \tag{S1}$$

and the hydrostatic primitive momentum prognostic equation:

$$\frac{\partial \mathbf{u}}{\partial t} = -\mathbf{u} \cdot \nabla_p \mathbf{u} - \omega \frac{\partial \mathbf{u}}{\partial p} - 2\Omega \sin\vartheta \, \mathbf{k} \times \mathbf{u} - \frac{u \tan\vartheta}{a} \mathbf{k} \times \mathbf{u} - \nabla_p \Phi + \mathbf{F_M} + \mathbf{F_T} + \mathbf{F_G}, \tag{S2}$$

where the variables are defined as follows: $\mathbf{u}$, horizontal velocity vector; $\omega$, vertical motion; $\Phi$, geopotential; $\Omega$, planetary rotation rate; $a$, planetary radius; $\vartheta$, latitude. The variables $\mathbf{F_M}$, $\mathbf{F_T}$, and $\mathbf{F_G}$ denote apparent momentum sources or sinks due to subgrid-scale moist convection, turbulence, and gravity wave drag. The variables $S_M$, $S_T$, and $S_C$ denote apparent moisture sources or sinks due to moist physics (including subgrid-scale moist convection, condensation, and evaporation of condensate, hereafter condensation), turbulence, and chemistry. In MERRA-2, tendencies due to dynamics refers to the combination of the first two terms on the right-hand side of Equation S1 (hereafter, $S_D$) and the first five terms on that of Equation S2 (hereafter, $\mathbf{F_D}$). Then, we derive the prognostic equation of $q^2$ by multiplying $2q$ to Equation S1 and applying the product rule ($2q \frac{\partial q}{\partial t} = \frac{\partial q^2}{\partial t}$):

$$\frac{\partial q^2}{\partial t} = -2q\mathbf{u} \cdot \nabla_p q - 2q\omega \frac{\partial q}{\partial p} + 2qS_M + 2qS_T + 2qS_C. \tag{S3}$$

Additionally, we derive the prognostic equation of $K$ by taking the dot product of $\mathbf{u}$ to Equation S2 and applying the product rule ($\mathbf{u} \cdot \frac{\partial \mathbf{u}}{\partial t} = \frac{\partial}{\partial t}\left(\frac{\mathbf{u} \cdot \mathbf{u}}{2}\right) = \frac{\partial K}{\partial t}$; $-\mathbf{u} \cdot (\mathbf{u} \cdot \nabla_p \mathbf{u}) = \mathbf{u} \cdot \nabla_p\left(\frac{\mathbf{u} \cdot \mathbf{u}}{2}\right) = -\mathbf{u} \cdot \nabla_p K$; $-\mathbf{u} \cdot \omega \frac{\partial \mathbf{u}}{\partial p} = -\omega \frac{\partial}{\partial p}\left(\frac{\mathbf{u} \cdot \mathbf{u}}{2}\right) = -\omega \frac{\partial K}{\partial p}$):

$$\frac{\partial K}{\partial t} = -\mathbf{u} \cdot \nabla_p K - \omega \frac{\partial K}{\partial p} - \mathbf{u} \cdot \nabla_p \Phi + \mathbf{u} \cdot \mathbf{F_M} + \mathbf{u} \cdot \mathbf{F_T} + \mathbf{u} \cdot \mathbf{F_G}, \tag{S4}$$

Adding $K$ times Equation S3 and $q^2$ times Equation S4 and applying the product rule ($K\frac{\partial q^2}{\partial t} + q^2\frac{\partial K}{\partial t} = \frac{\partial q^2 K}{\partial t}$) yields the VKE prognostic equation:

$$\frac{\partial q^2 K}{\partial t} = -q^2 \mathbf{u} \cdot \boldsymbol{\nabla}_p K - q^2 \omega \frac{\partial K}{\partial p} - q^2 \mathbf{u} \cdot \boldsymbol{\nabla}_p \Phi + q^2 \mathbf{u} \cdot \mathbf{F_M} + q^2 \mathbf{u} \cdot \mathbf{F_T} + q^2 \mathbf{u} \cdot \mathbf{F_G}$$

$$-2Kq\mathbf{u} \cdot \boldsymbol{\nabla}_p q - 2Kq\omega \frac{\partial q}{\partial p} + 2KqS_M + 2KqS_T + 2KqS_C, \quad (S5)$$

Taking the time derivative of Equation 2 and using the Leibniz integral rule yields a simple form of IVKE prognostic equation:

$$\frac{\partial}{\partial t}\langle q^2 K \rangle = \langle \frac{\partial q^2 K}{\partial t} \rangle + \left[\frac{q^2 K}{g}\right]_{p_B} \frac{\partial p_B}{\partial t}, \quad (S6)$$

where $\langle \ \rangle$ denotes the following integral operator:

$$\langle \ \rangle \equiv -\frac{1}{g}\int_{p_B}^{p_T}(\ )\mathrm{d}p.$$

Plugging Equation S5 into S6 yields the IVKE prognostic equation with complete physics:

$$\frac{\partial}{\partial t}\langle q^2 K \rangle = \underbrace{\langle -q^2 \mathbf{u} \cdot \boldsymbol{\nabla}_p K \rangle}_{\text{HAKE}} + \underbrace{\langle -q^2 \omega \frac{\partial K}{\partial p} \rangle}_{\text{VAKE}} + \underbrace{\langle -q^2 \mathbf{u} \cdot \boldsymbol{\nabla}_p \Phi \rangle}_{\text{PEKE}} + \underbrace{\langle q^2 \mathbf{u} \cdot \mathbf{F_M} \rangle}_{\text{MOKE}} + \underbrace{\langle q^2 \mathbf{u} \cdot \mathbf{F_T} \rangle}_{\text{TOKE}} + \underbrace{\langle q^2 \mathbf{u} \cdot \mathbf{F_G} \rangle}_{\text{GOKE}}$$

$$+\underbrace{\langle -2Kq\mathbf{u} \cdot \boldsymbol{\nabla}_p q \rangle}_{\text{HAV}} + \underbrace{\langle -2Kq\omega \frac{\partial q}{\partial p} \rangle}_{\text{VAV}} + \underbrace{\langle 2KqS_M \rangle}_{\text{COV}} + \underbrace{\langle 2KqS_T \rangle}_{\text{TOV}} + \underbrace{\langle 2KqS_C \rangle}_{\text{CMOV}}$$

$$+ \left[\frac{q^2 K}{g}\right]_{p_B} \frac{\partial p_B}{\partial t}, \quad (S7)$$

The acronyms are defined in Table S1. Moreover, the MERRA-2 provides nonphysical tendencies due to incorporating analysis data into winds ($\mathbf{F_A}$) and moisture ($S_A$), so we also calculate an analysis effect added to the right-hand side of Equation S7 as $\langle q^2 \mathbf{u} \cdot \mathbf{F_A} + 2KqS_A \rangle$.

*Calculating the IVKE budget using MERRA-2 and ERA5 datasets*

Here, we detail the methodology for assessing uncertainties in the IVKE budget using MERRA-2 data. We estimate the uncertainty of HAKE, VAKE, and PEKE using the residual of the dynamic tendency of KE ($R_{DOKE}$):

$$R_{DOKE} = \langle q^2 \mathbf{u} \cdot \mathbf{F_D} \rangle - \left( \langle -q^2 \mathbf{u} \cdot \nabla_p K \rangle + \langle -q^2 \omega \frac{\partial K}{\partial p} \rangle + \langle -q^2 \mathbf{u} \cdot \nabla_p \Phi \rangle \right). \tag{S8}$$

We estimate the uncertainty of HAV and VAV using the residual of the dynamic tendency of vapor ($R_{DOV}$):

$$R_{DOV} = \langle 2Kq S_D \rangle - \left( \langle -2Kq \mathbf{u} \cdot \nabla_p q \rangle + \langle -2Kq \omega \frac{\partial q}{\partial p} \rangle \right). \tag{S9}$$

We estimate the uncertainty of all the other terms using the residual of the following equation ($R_{total}$):

$$R_{total} = \langle \frac{\partial q^2 K}{\partial t} \rangle - (\langle q^2 \mathbf{u} \cdot \mathbf{F_D} \rangle + \langle q^2 \mathbf{u} \cdot \mathbf{F_M} \rangle + \langle q^2 \mathbf{u} \cdot \mathbf{F_T} \rangle + \langle q^2 \mathbf{u} \cdot \mathbf{F_G} \rangle$$
$$+ \langle 2Kq S_D \rangle + \langle 2Kq S_M \rangle + \langle 2Kq S_T \rangle + \langle 2Kq S_C \rangle + \langle q^2 \mathbf{u} \cdot \mathbf{F_A} + 2Kq S_A \rangle). \tag{S10}$$

ERA5 does not provide $\mathbf{F_M}$, $\mathbf{F_T}$, $\mathbf{F_G}$, $S_M$, $S_T$, and $S_C$. However, MERRA-2 results suggest that $\mathbf{F_T}$ and $S_M$ dominate. Thus, from Equation S3 and S4, we estimate them as follows:

$$\langle q^2 \mathbf{u} \cdot \mathbf{F_T} \rangle = q^2 \frac{\partial}{\partial t} \langle K \rangle - \langle -q^2 \mathbf{u} \cdot \nabla_p K \rangle - \langle -q^2 \omega \frac{\partial K}{\partial p} \rangle - \langle -q^2 \mathbf{u} \cdot \nabla_p \Phi \rangle, \tag{S11}$$

$$\langle 2Kq S_M \rangle = K \frac{\partial}{\partial t} \langle q^2 \rangle - \langle -2Kq \mathbf{u} \cdot \nabla_p q \rangle - \langle -2Kq \omega \frac{\partial q}{\partial p} \rangle. \tag{S12}$$

*PE-to-KE conversion decomposition*

The PEKE term can be decomposed into two physical processes. First, use the product rule:

$$\langle -q^2 \mathbf{u} \cdot \mathbf{\nabla}_p \Phi \rangle = \langle q^2 \Phi \mathbf{\nabla}_p \cdot \mathbf{u} \rangle + \langle -q^2 \mathbf{\nabla}_p \cdot (\mathbf{u}\Phi) \rangle.$$

Then, we use the continuity equation:

$$\langle -q^2 \mathbf{u} \cdot \mathbf{\nabla}_p \Phi \rangle = \langle -q^2 \Phi \frac{\partial \omega}{\partial p} \rangle + \langle -q^2 \mathbf{\nabla}_p \cdot (\mathbf{u}\Phi) \rangle.$$

Then, we use the product rule again:

$$\langle -q^2 \mathbf{u} \cdot \mathbf{\nabla}_p \Phi \rangle = \langle q^2 \omega \frac{\partial \Phi}{\partial p} \rangle + \langle -q^2 \frac{\partial}{\partial p}(\omega \Phi) \rangle + \langle -q^2 \mathbf{\nabla}_p \cdot (\mathbf{u}\Phi) \rangle.$$

Last, we use the hydrostatic equation:

$$\langle -q^2 \mathbf{u} \cdot \mathbf{\nabla}_p \Phi \rangle = \langle -q^2 \omega \alpha \rangle + \langle -q^2 \frac{\partial}{\partial p}(\omega \Phi) - q^2 \mathbf{\nabla}_p \cdot (\mathbf{u}\Phi) \rangle, \tag{S13}$$

where $\alpha$ denotes specific volume. The two brackets on the right-hand side of Equation S13 are baroclinic conversion and geopotential flux convergence. These terms largely offset each other in the composite AR (Figure S15), such that the PEKE term accounts for roughly 3% of the baroclinic conversion in terms of their contributions to the composite AR growth.

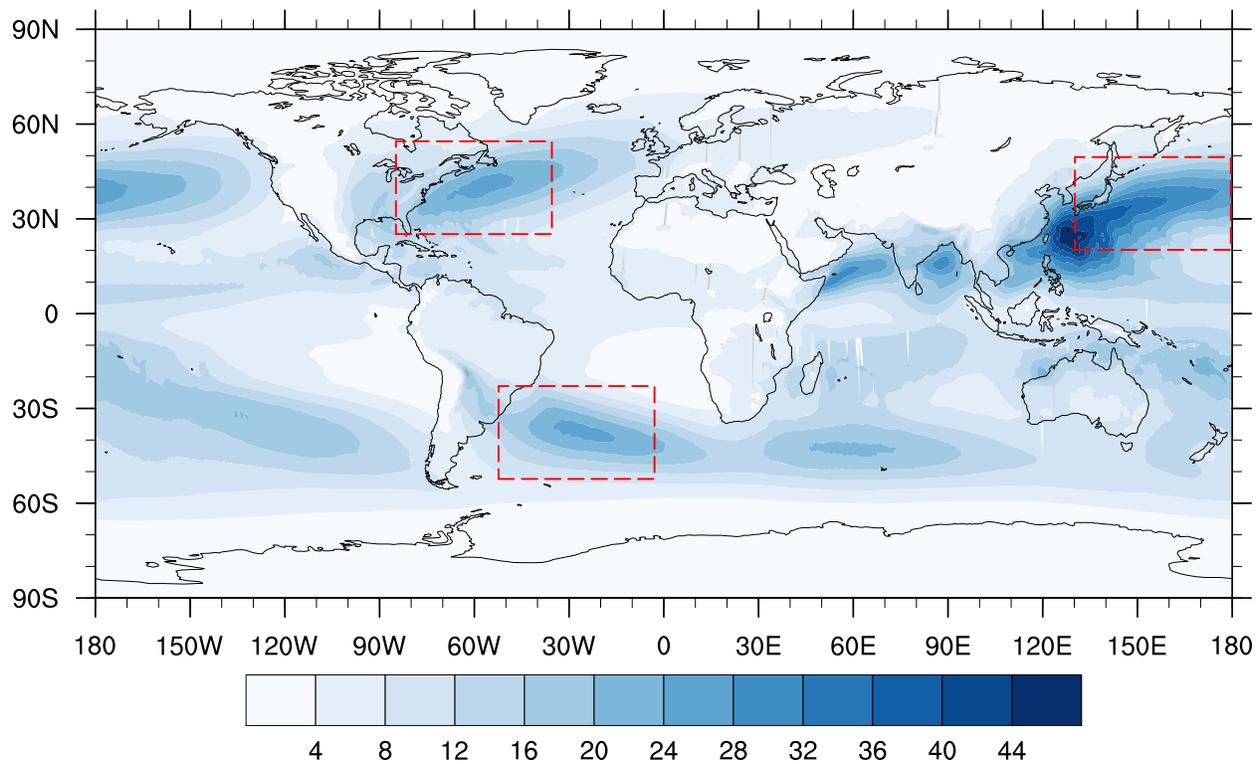

**Figure S1.** Temporal sample standard deviation of IVKE (units: kg s$^{-2}$) from 1980 to 2019 from MERRA-2 data. The red dash box highlights the active AR regions we selected.

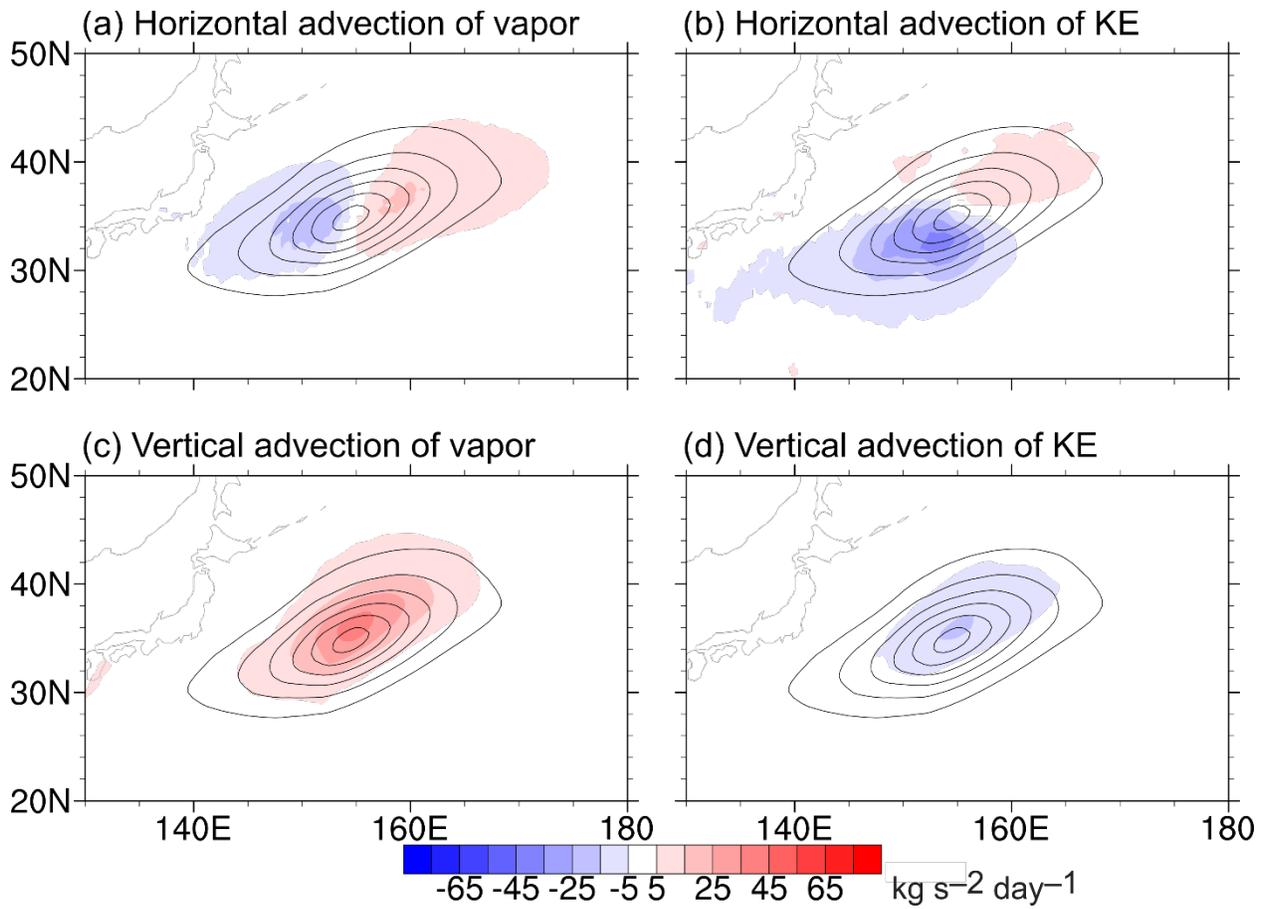

**Figure S2.** Like Figure 2 but with further decomposition of the shaded fields in Figure 2b and 2d. The shading denotes IVKE tendencies due to (a) horizontal advection of vapor, (b) horizontal advection of KE, (c) vertical advection of vapor, and (d) vertical advection of KE.

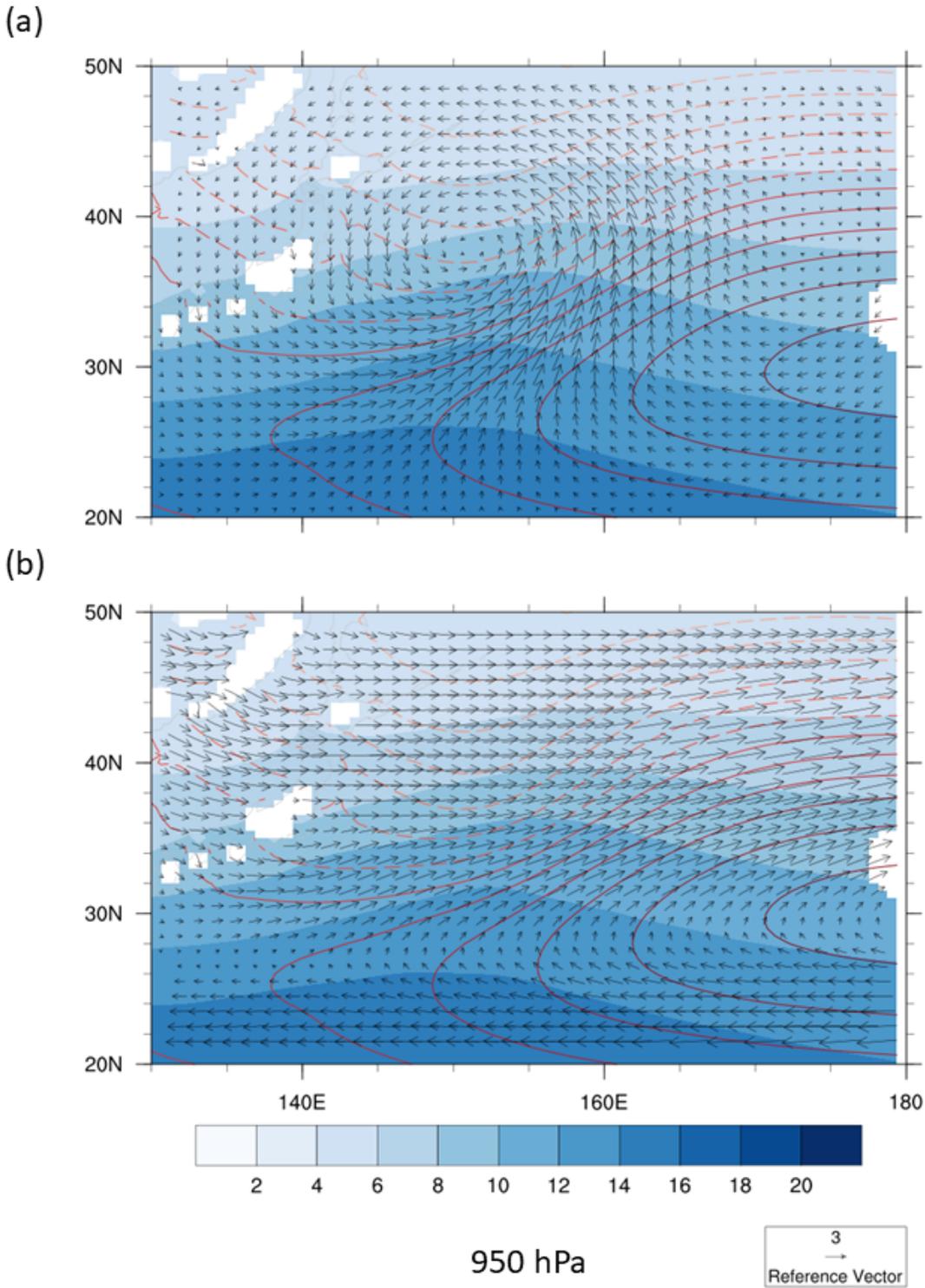

**Figure S3.** Total geopotential height (contours, interval: 10 m), total specific humidity (shading, units: g kg$^{-1}$), and (a) anomalous winds (arrows, units: m s$^{-1}$) (b) temporal mean winds at 950

hPa of the composite AR in Figure 2. Solid reddish contours are higher than the areal mean geopotential height in the plotting domain, and dashed yellowish contours are lower than that. The AR composite anomalies are calculated by regressing upon areal mean IVKE in the 1°×1° box centered at 155°E 35°N from MERRA-2 data from 2010 to 2019. The total geopotential height is calculated by adding the AR composite geopotential height anomaly and the temporal-mean geopotential height.

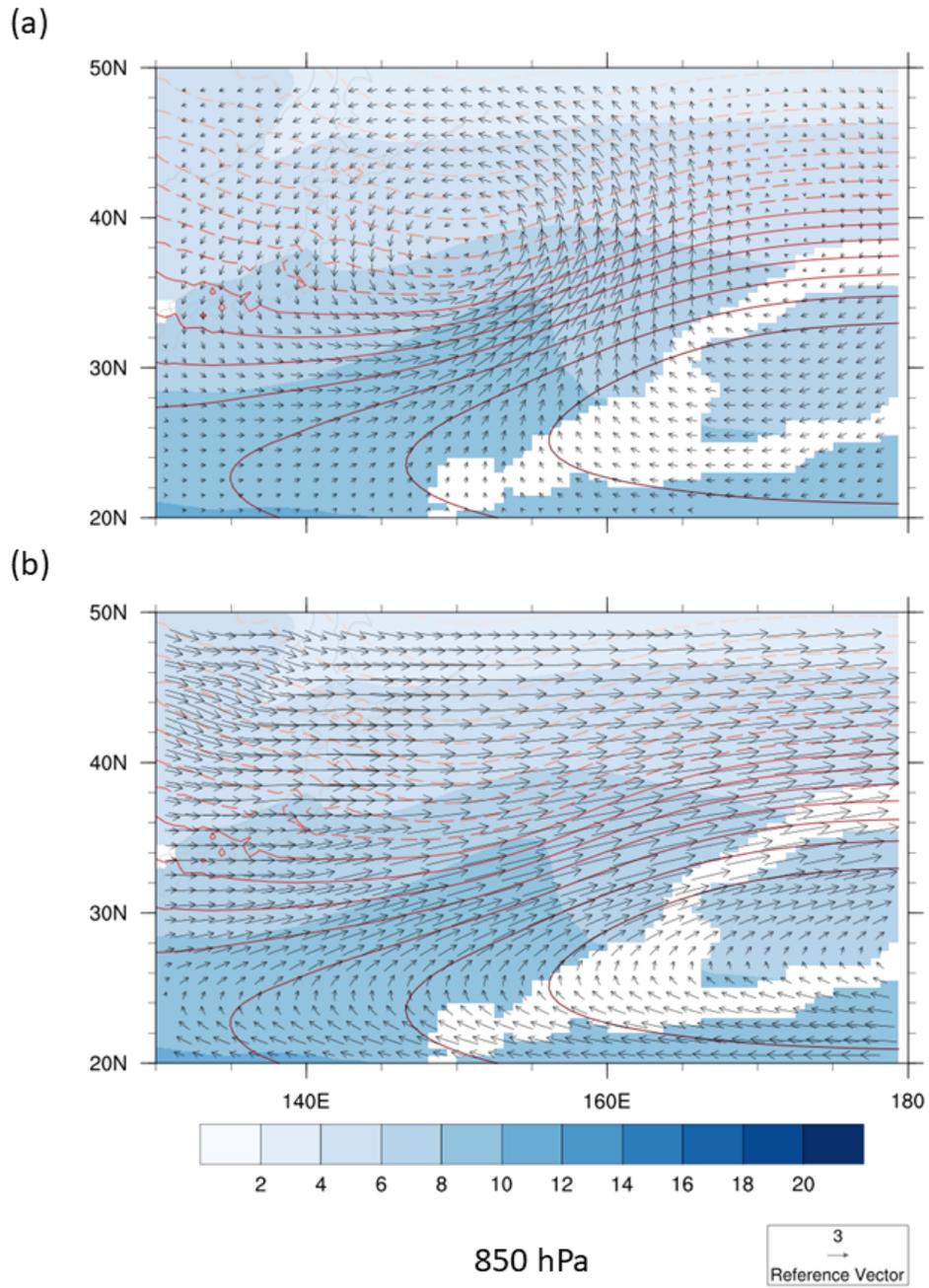

**Figure S4.** Like Figure S3 but at 850 hPa.

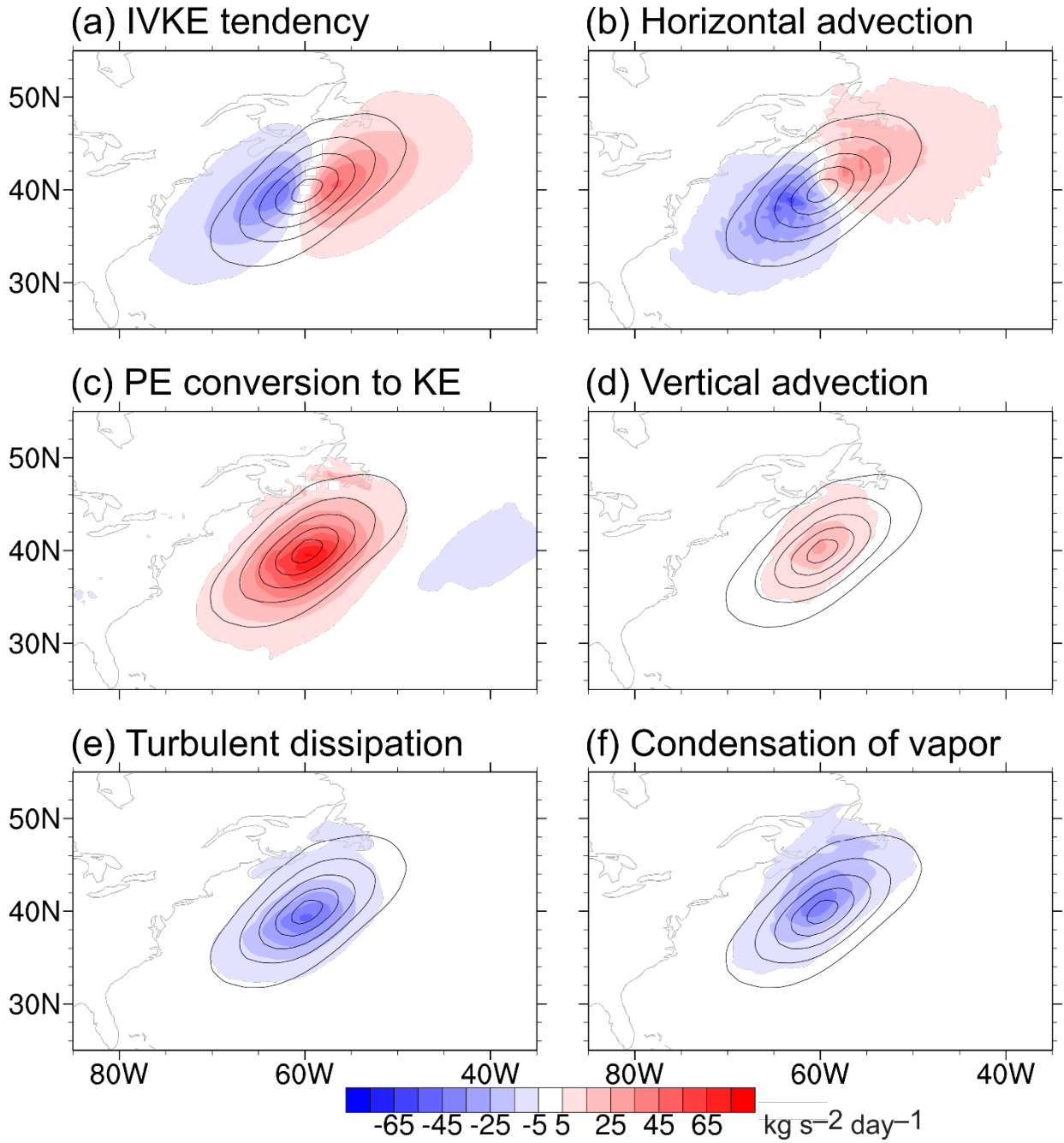

**Figure S5.** Like Figure 2 but regressing upon areal mean IVKE in the 1°×1° box centered at 60°W 40°N.

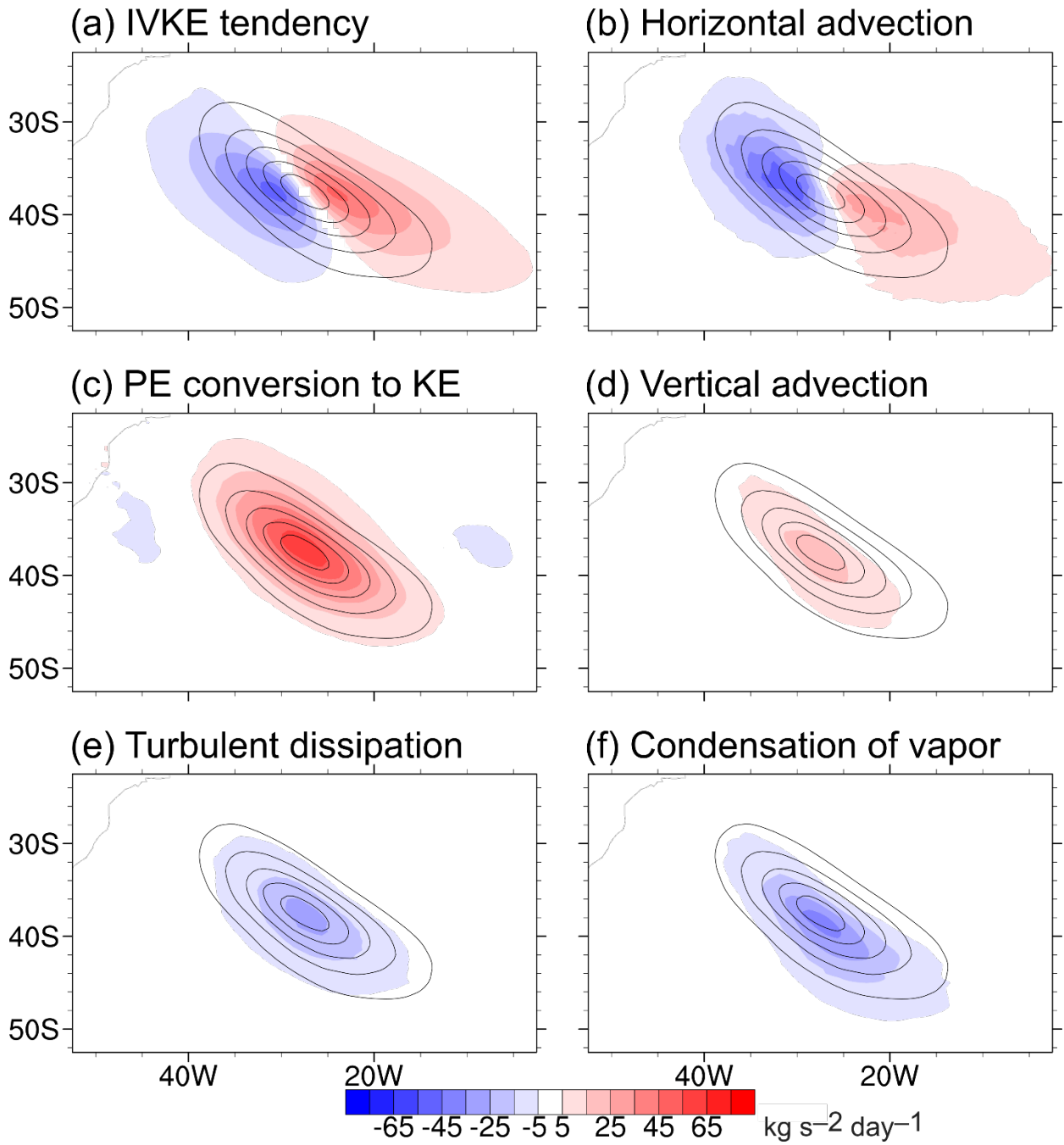

**Figure S6.** Like Figure 2 but regressing upon areal mean IVKE in the 1°×1° box centered at 27.5°W 37.5°S.

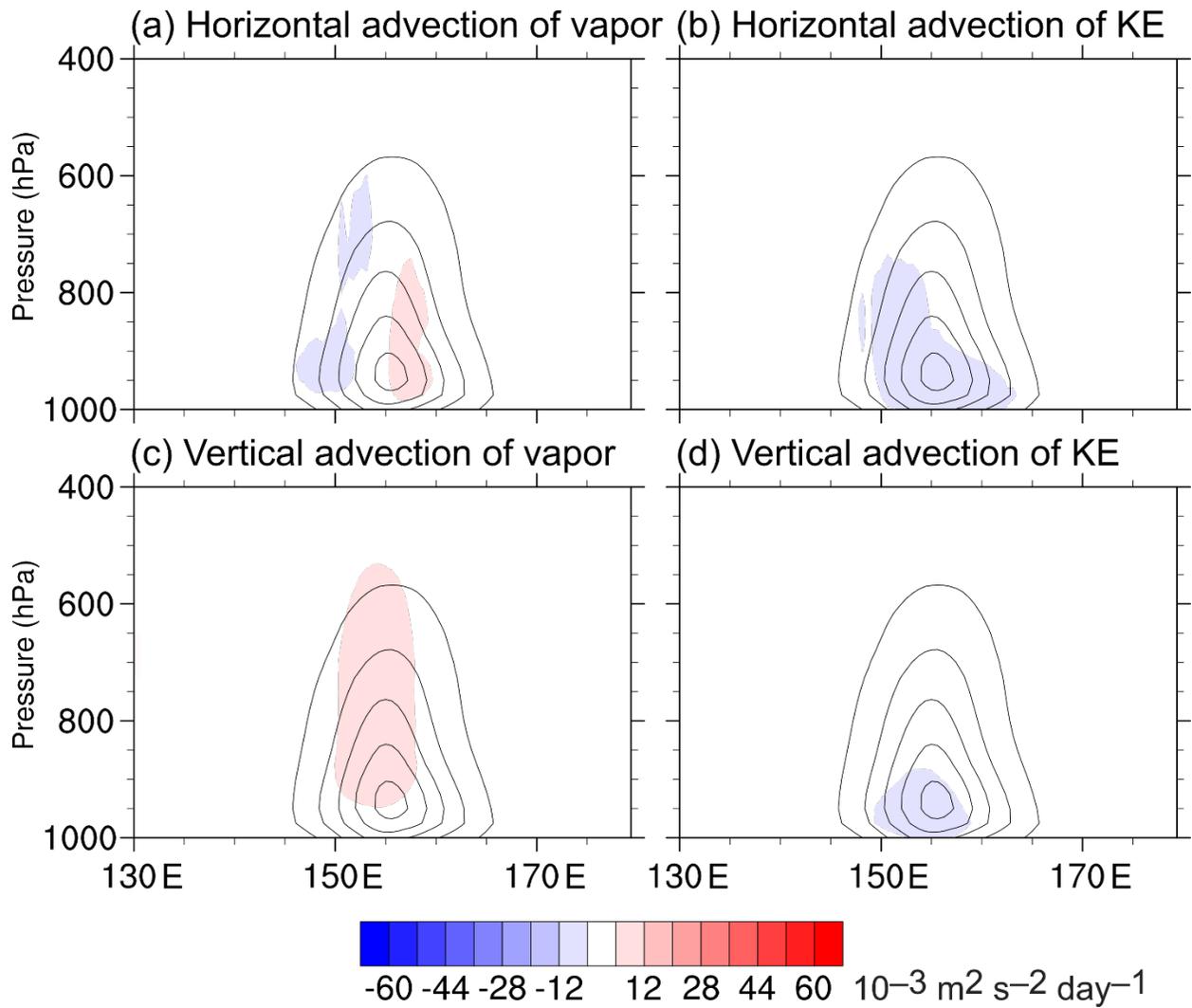

**Figure S7.** Like Figure 3 but with further decomposition of the shaded fields in Figure 3b and 3d. The shading (units: $10^{-3}$ m² s⁻² day⁻¹) denotes VKE tendencies due to (a) horizontal advection of vapor, (b) horizontal advection of KE, (c) vertical advection of vapor, and (d) vertical advection of KE.

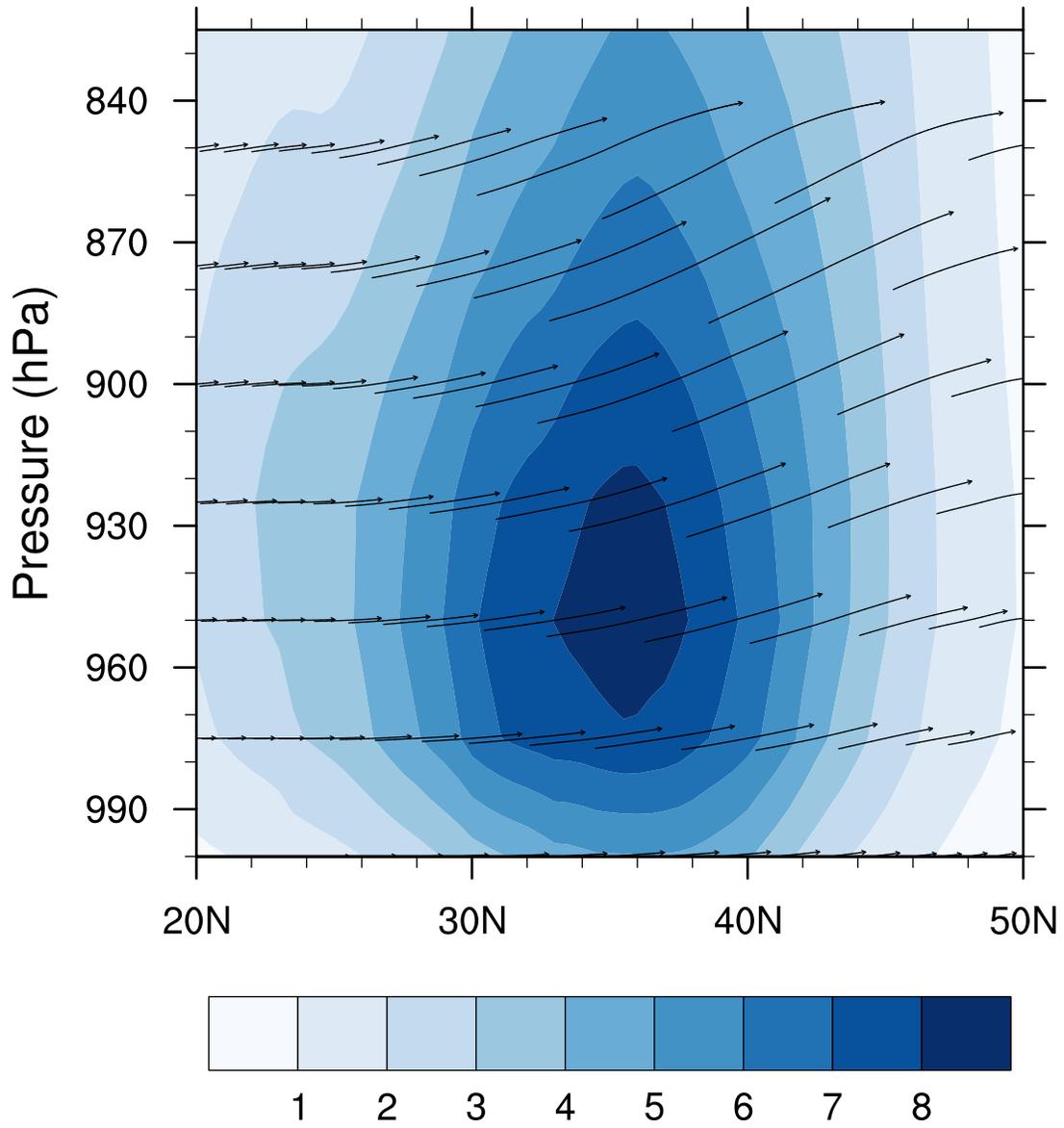

**Figure S8.** Like Figure 3 but vertical cross section along 155°E of anomalous VKE (shades, units: $10^{-3}$ m$^2$ s$^{-2}$) and anomalous air motion (arrows) of the AR composite (average taken between 154.5°E and 155.5°E).

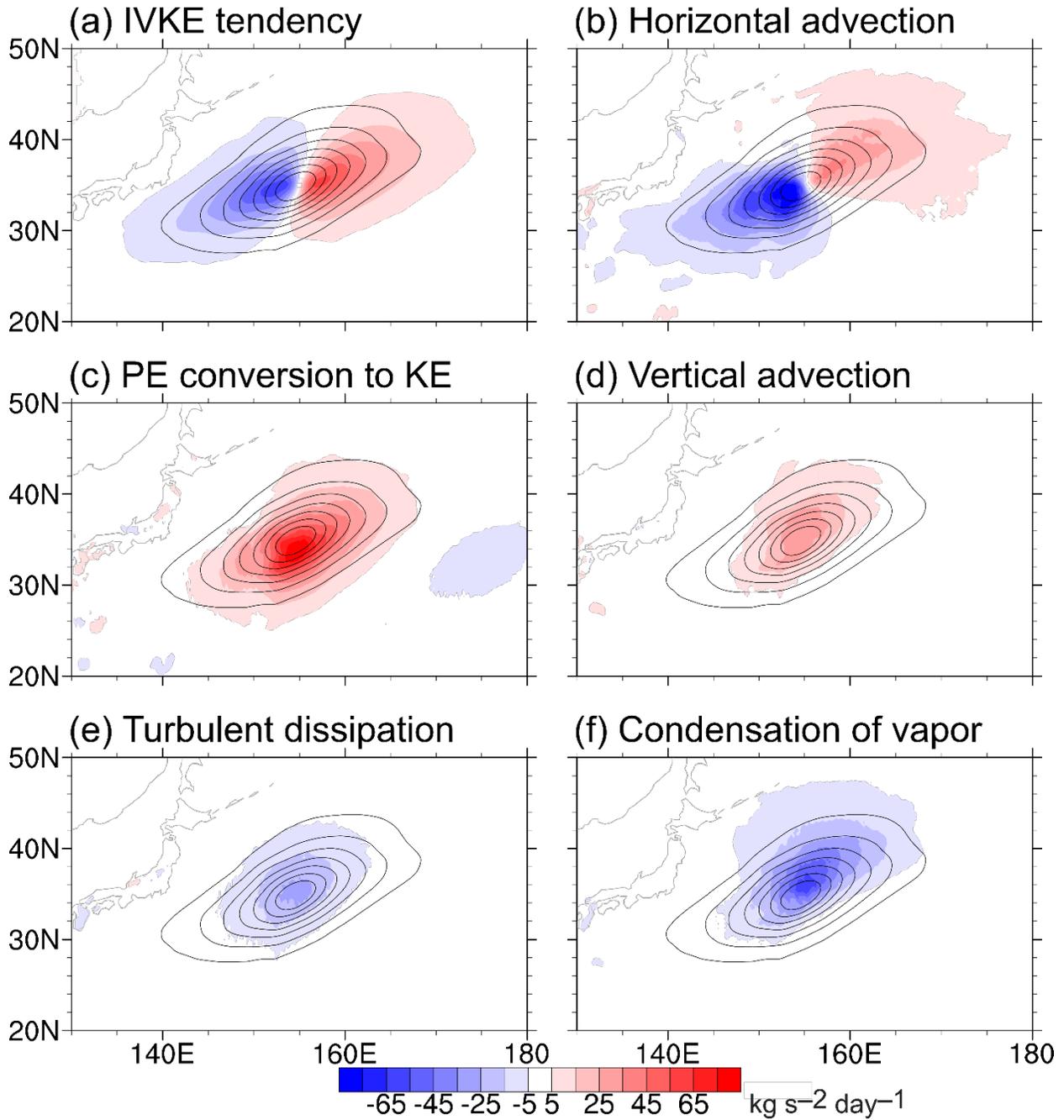

**Figure S9.** Like Figure 2 but using ERA5 data. The KE and vapor parts of the residuals of ERA5 analysis are taken as (e) turbulent dissipation of KE and (f) condensation of vapor, respectively
15__________________

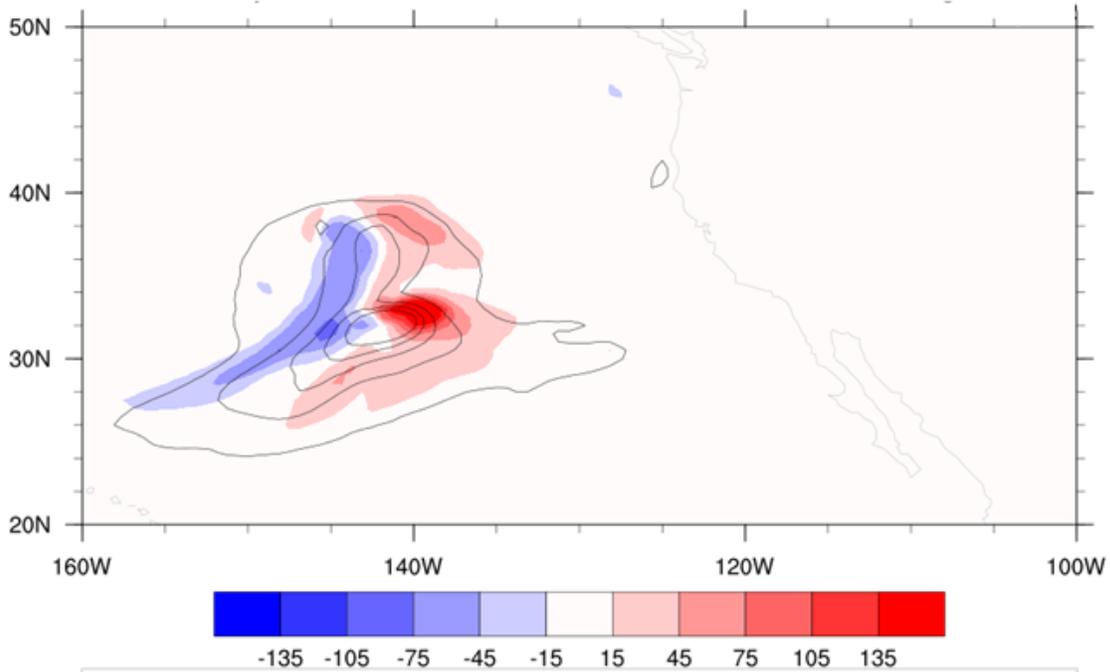

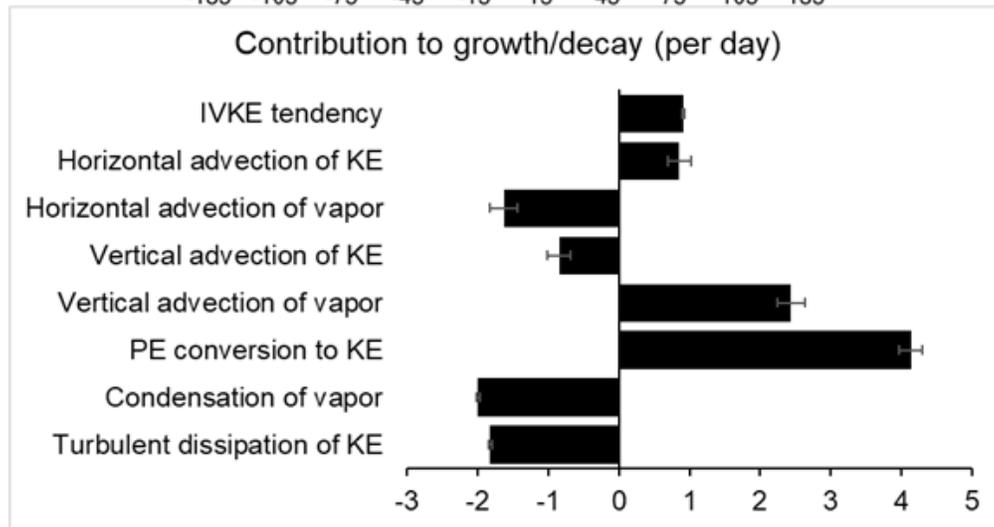

**Figure S10.** (a) Temporal-mean IVKE (contours, minimum level: 15 kg s$^{-2}$, interval: 30 kg s$^{-2}$) and IVKE tendency (shading, units: kg s$^{-2}$ per 3 hours) during 18 to 21 UTC, 3 January 2023 (18 hours before the AR landfall onto the west coast of North America, intensification stage) from MERRA-2 data. Panel (b) is like Figure 4a but using temporal-mean data from this three-hour period.

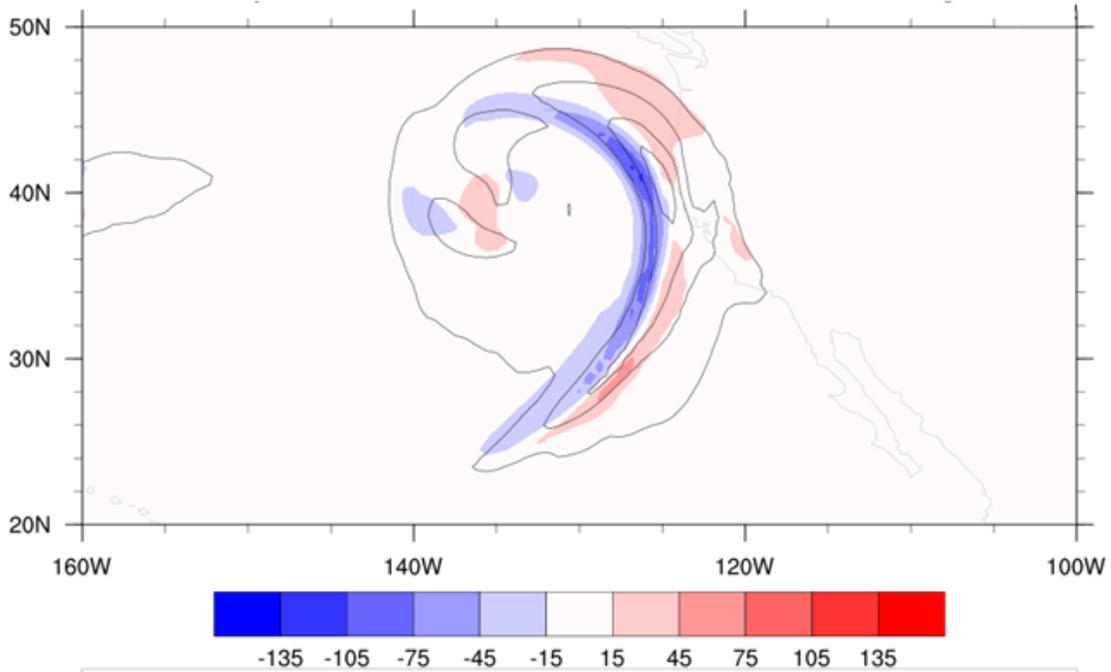

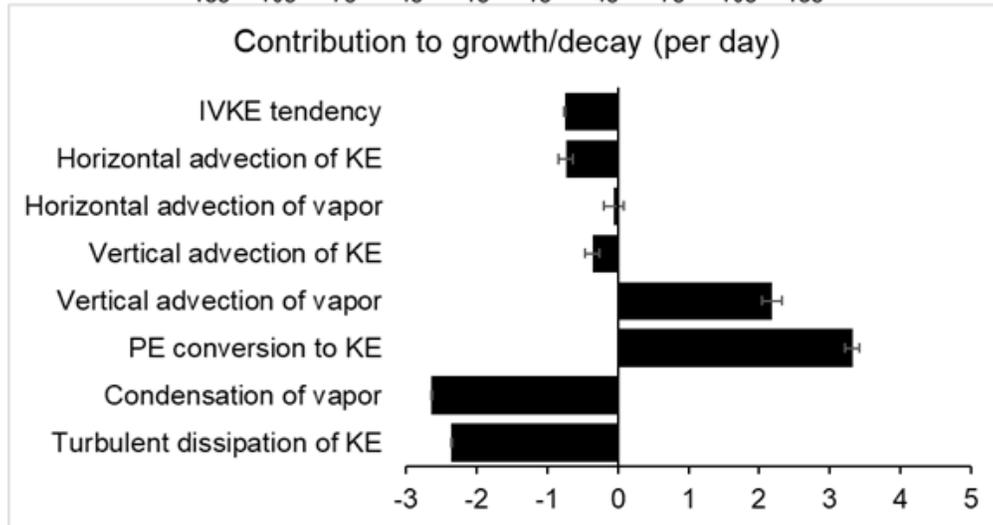

**Figure S11.** Like Figure S8 but for 18 to 21 UTC, 4 January 2023 (6 hours after the AR landfall onto the west coast of North America, decaying stage).

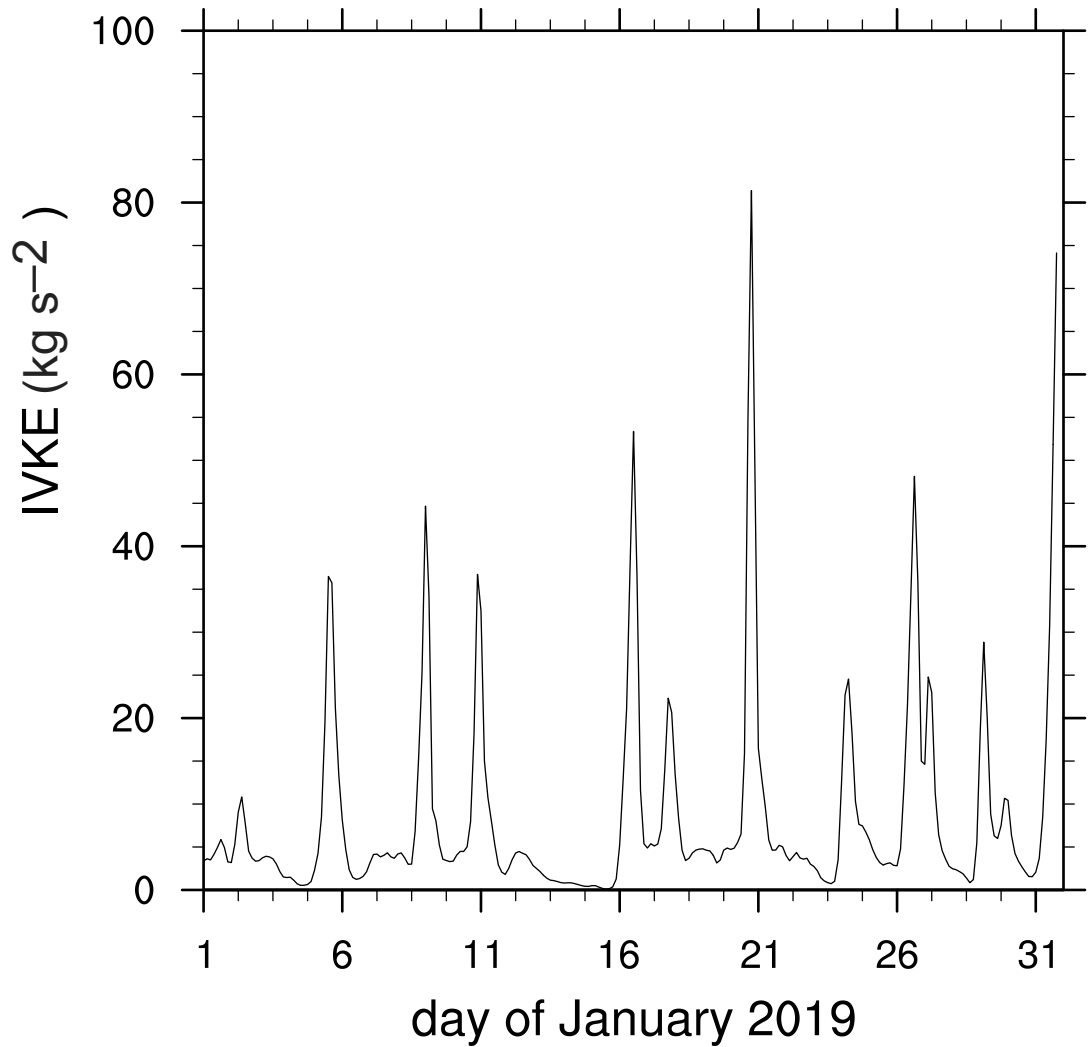

**Figure S12.** Three-hourly time series of IVKE in the 1°×1° box centered at 155°E 35°N in January 2019 from MERRA-2 data.

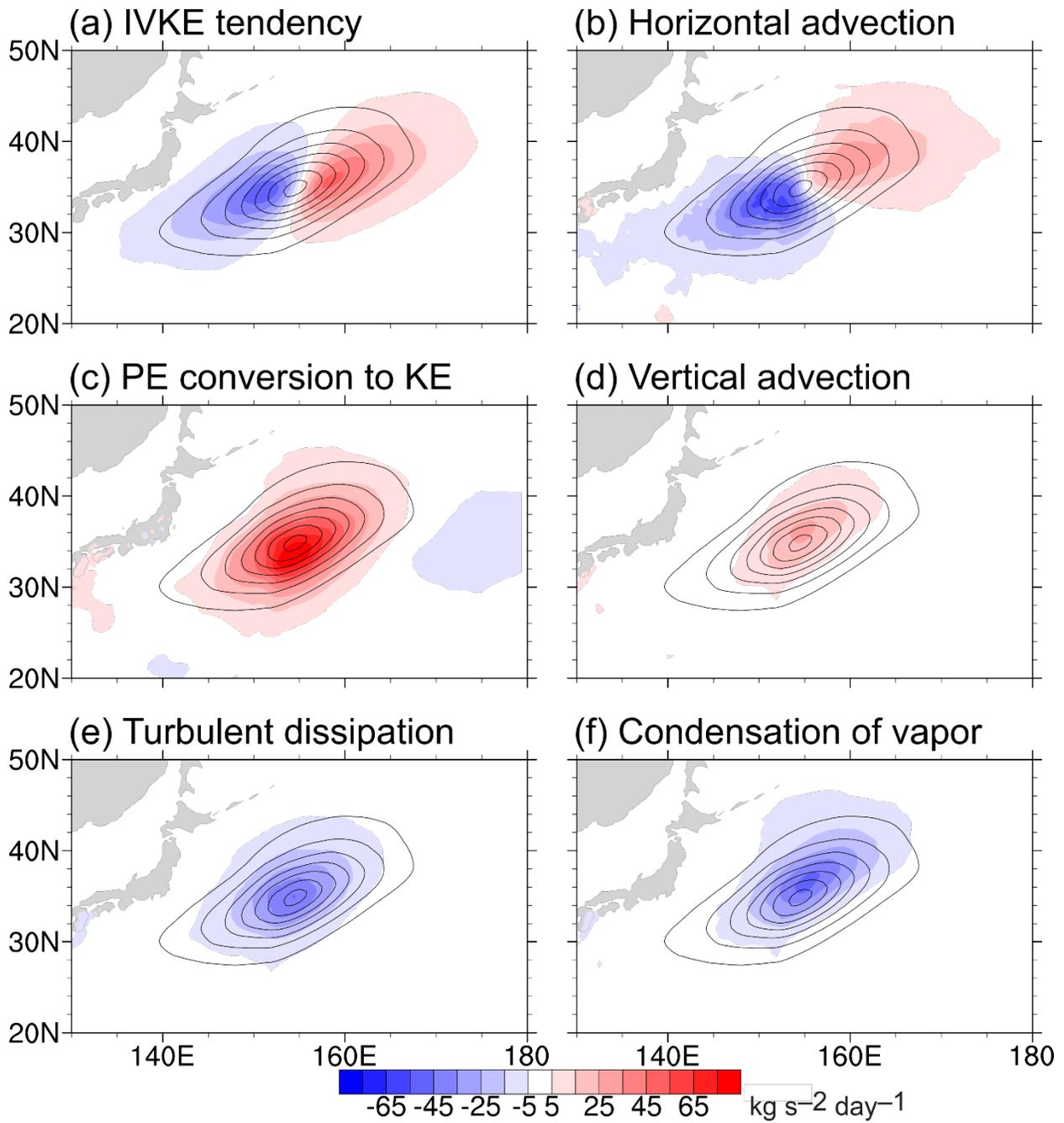

**Figure S13.** Like Figure 2 but excluding the time steps when AR is not detected in the 1°×1° box centered at 155°E 35°N using the TempestExtremes.

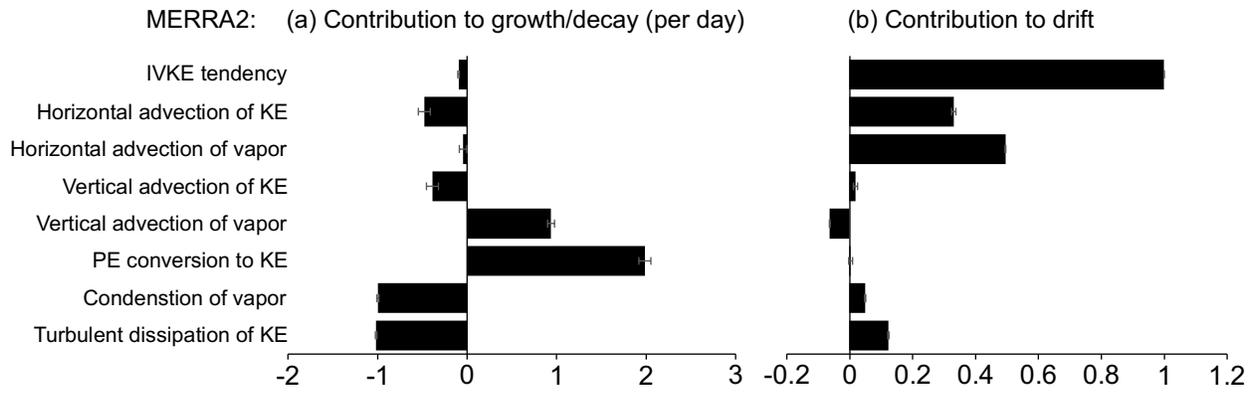

**Figure S14.** Like Figure 4ab but the region of projection is within 135°E to 175°, 25°N to 45°N.

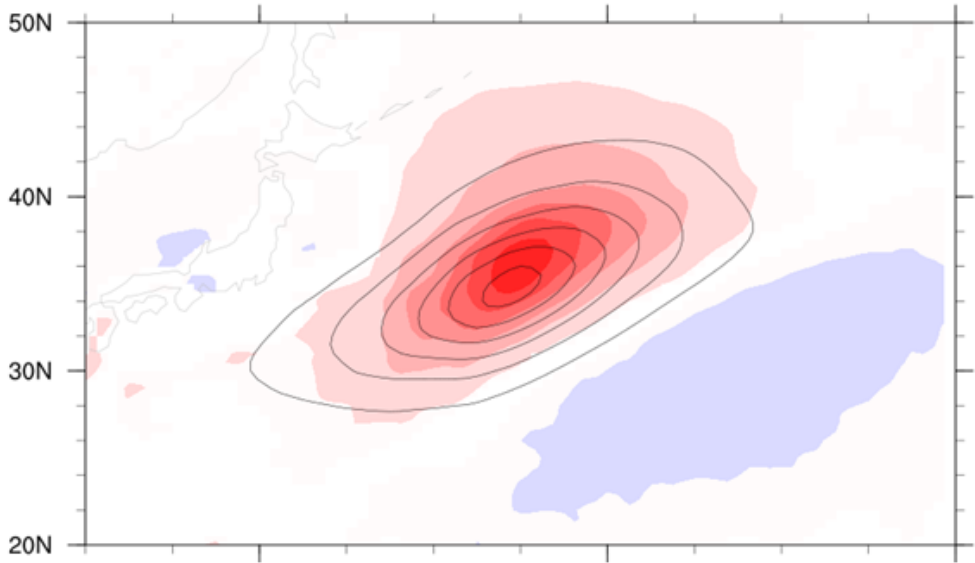
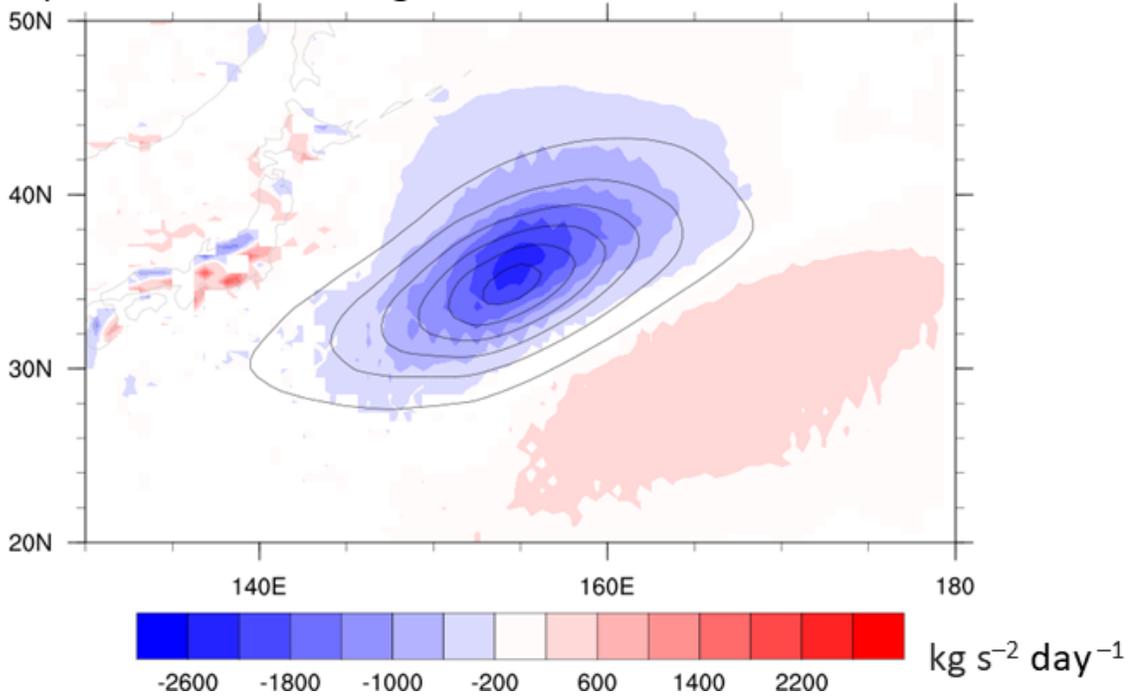

**Figure S15.** Like Figure 2 but with further decomposition of the PEKE term (the shaded fields in Figure 2c). See Equation S13. The shading denotes IVKE tendencies due to (a) baroclinic conversion, (b) geopotential flux convergence. The color scale is different from Figure 2.

**Table S1.** Contribution of physical processes to the composite AR in Figure 2 using MERRA-2.

| | Growth or decay (%)[1] | Drift (%)[2] |
|---|---|---|
| IVKE tendency | -3.1 | 100.0 |
| Horizontal advection of KE (HAKE) | -15.9 | +33.6 |
| Vertical advection of KE (VAKE) | -12.8 | +2.0 |
| Potential energy conversion to KE (PEKE) | +65.8 | +0.1 |
| Turbulent dissipation of KE (TOKE) | -33.7 | 12.3 |
| Moist convection tendency of KE (MOKE) | +0.7 | -0.6 |
| Gravity wave drag of KE (GOKE) | 0.0 | 0.0 |
| Horizontal advection of vapor (HAV) | -1.4 | +49.7 |
| Vertical advection of vapor (VAV) | +31.1 | -6.3 |
| Condensation of vapor (COV) | -33.1 | +5.0 |
| Turbulence tendency of vapor (TOV) | +2.3 | -4.1 |
| Chemistry tendency of vapor (CMOV) | 0.0 | 0.0 |
| Surface pressure tendency effect | -0.6 | -1.1 |
| Analysis effect | +2.2 | +2.0 |
| Residual (DOKE, see Eq. S8) | -4.5 | +1.4 |
| Residual (DOV, see Eq. S9) | -2.7 | +0.2 |
| Residual (total, see Eq. S10) | +0.6 | +0.4 |

---

[1] Percentage to the sum of all the physical sources (*i.e.*, positive terms except analysis effect and residual)
[2] Percentage to the IVKE tendency